# Assessment of Data Consistency through Cascades of Independently Recurrent Inference Machines for fast and robust accelerated MRI reconstruction

D. Karkalousos[1], S. Noteboom[2], H. E. Hulst[2,3], F.M. Vos[4], M.W.A. Caan[1]

[1] Department of Biomedical Engineering & Physics, Amsterdam UMC, University of Amsterdam, Amsterdam, Netherlands

[2] Department of Anatomy & Neurosciences, MS Center Amsterdam, Amsterdam Neuroscience, Amsterdam UMC, Vrije Universiteit Amsterdam, Amsterdam, Netherlands

[3] Department of Medical, Health and Neuropsychology, Leiden University, Leiden, Netherlands

[4] Department of Imaging Physics, Delft University of Technology, Delft, Netherlands

E-mail: d.karkalousos@amsterdamumc.nl

## Abstract

**Objective:** Machine Learning methods can learn how to reconstruct Magnetic Resonance Images (MRI) and thereby accelerate acquisition, which is of paramount importance to the clinical workflow. Physics-informed networks incorporate the forward model of accelerated MRI reconstruction in the learning process. With increasing network complexity, robustness is not ensured when reconstructing data unseen during training. We aim to embed data consistency (DC) in deep networks while balancing the degree of network complexity. While doing so, we will assess whether either explicit or implicit enforcement of DC in varying network architectures is preferred to optimize performance.

**Approach:** We propose a scheme called Cascades of Independently Recurrent Inference Machines (CIRIM) to assess DC through unrolled optimization. Herein we assess DC both implicitly by gradient descent and explicitly by a designed term. Extensive comparison of the CIRIM to Compressed Sensing as well as other Machine Learning methods is performed: the End-to-End Variational Network (E2EVN), CascadeNet, KIKINet, LPDNet, RIM, IRIM, and UNet. Models were trained and evaluated on $T_1$-weighted and FLAIR contrast brain data, and $T_2$-weighted knee data. Both 1D and 2D undersampling patterns were evaluated. Robustness was tested by reconstructing 7.5x prospectively undersampled 3D FLAIR MRI data of Multiple Sclerosis (MS) patients with white matter lesions.

**Main results:** The CIRIM performed best when implicitly enforcing DC, while the E2EVN required an explicit DC formulation. Through its cascades, the CIRIM was able to score higher on Structural Similarity and PSNR compared to other methods, in particular under heterogeneous imaging conditions. In reconstructing MS patient data, prospectively acquired with a sampling pattern unseen during model training, the CIRIM maintained lesion contrast while efficiently denoising the images.

**Significance:** The CIRIM showed highly promising generalization capabilities maintaining a very fair trade-off between reconstructed image quality and fast reconstruction times, which is crucial in the clinical workflow.

## 1. Introduction

Magnetic Resonance Imaging (MRI) non-invasively images the anatomy of the human body. It is important to note that data are acquired in the frequency domain, known as k-space. Conventionally, the measured signals need to adhere to the Nyquist-criterion to allow for inverse Fourier transforming them to the image domain without aliasing. Due to hardware limitations and physical constraints, however, sampling the full k-space leads to long scanning times. Almost 25 years ago, Parallel-Imaging (PI) [1] was introduced to reduce acquisition times, overcoming hardware and software limitations by applying multiple receiver coil arrays. Each coil has a distinct sensitivity profile which can be exploited in reconstructing undersampled data. With sensitivity encoding (SENSE) the multicoil data are transformed to the image domain through the inverse Fourier Transform, after which a reconstruction algorithm dealiases the images based on the coil sensitivity maps [2]. The combination of PI with Compressed Sensing (CS) [3,4] is now standardly applied in clinical settings, allowing for high acceleration factors by utilizing the constrained reconstruction through a sparsifying transform.

Machine Learning (ML) methods can learn how to reconstruct images by training them on acquired data for which a reference reconstruction is available. As such the reconstruction times can be reduced, which is of paramount importance to the clinical workflow. The UNet [5] may be the most popular network in the field and the base for numerous other methods, as elaborated upon below. Its unique architecture, with the down- and up- sampling operators and the large number of features on the output, has made it a cornerstone approach in image reconstruction today. Although such a network architecture can perform well, its performance is still limited due to operating only in image space without any MR physics knowledge incorporated.

Physics-informed networks were therefore introduced, incorporating the forward model of accelerated MRI reconstruction in the learning process. The Variational Network (VN) [6] and the Recurrent Inference Machines (RIM) [7–9] proposed to solve the inverse problem of accelerated MRI reconstruction through a Bayesian estimation. Alternatively, scan-specific techniques were used to restore missing k-space from fully-sampled autocalibration data [10–12]. Furthermore, dual-domain networks were proposed to leverage the k-space information and perform corrections both in the frequency domain and the image domain. The Learned Primal-Dual reconstruction technique (LPDNet) [13] replaced the proximal operators in the Primal-Dual Hybrid Gradient algorithm [14] with learned operators, yielding a learning scheme combined with model-based reconstruction. The KIKI-net [15] introduced a sequence of Convolutional Neural Networks (CNN) performed in k-space (K) and image space (I). Later, concatenations of UNets were applied to replace the sequence of CNNs in the KIKI-net [16]. Finally, the Model-Based Deep Learning technique [17] proposed a learned model-based reconstruction scheme involving a data consistency term.

With increasing network complexity, however, robustness is not ensured when reconstructing data unseen during training. This especially concerns clinical data with pathology for which fully sampled reference data cannot be obtained. This was understood in recent MRI reconstruction challenges [18–20], in which deep end-to-end schemes, such as the End-to-End Variational Network (E2EVN) [21], the XPDNet [22], and the Joint-ICNet [23] allowed for higher image quality at increased acceleration factors but not necessarily for generalization to out-of-distribution data containing pathologies. Recurrent Neural Networks (RNNs), i.e., the RIM and the Pyramid Convolutional RNN [24], appeared to generalize well on out-of-distribution data due to their nature of maintaining a notion of memory [25]. However, they scored lower on the trained data compared to the previously mentioned networks, possibly due to a limited number of iterations required to avoid gradient instabilities. Such methods would potentially benefit of increased network complexity as can be achieved using a number of cascades of networks [26–28]. The cascades can be considered as stacked networks targeting to resolve aliasing artifacts and to enhance denoising by iteratively evaluating the reconstruction, but without sharing parameters through backpropagation. Unfortunately, a solution may no longer be consistent with the acquired data with increasing network complexity. This raises a need for embedding data consistency in deep networks while balancing the degree of network complexity.

Data Consistency (DC) can be embedded into the learning scheme in several ways, such as through gradient descent [6,7,21,29], an iterative energy minimization process, namely variable-splitting [30], Generative Adversarial Networks [31–34], adversarial transformers [35], complex-valued networks [36–38], transfer learning [39], manifold approximation [40], or through sparsity [41–45]. Recent work evaluated enforcing DC in three ways, by gradient descent, by proximal mapping, and by variable-splitting [46]. It was shown that the training set could be reduced in size by doing so. The best results were obtained when train and test domains were aligned. However, it remains unknown whether either explicit or implicit enforcement of DC in varying network architectures is the best approach to optimize performance.

This work proposes a scheme called Cascades of Independently Recurrent Inference Machines (CIRIM). The CIRIM comprise RIM blocks sequentially connected through cascades and the efficient Independently Recurrent Neural Network (IndRNN) [47] as recurrent unit. The cascades allow us to train a deep but balanced RNN for improved dealiasing and denoising, while maintaining stable gradient calculations. The enforcement of DC in an implicit or explicit manner will be assessed by comparison to the E2EVN. The networks are further compared to the CascadeNet [26], the KIKINet [15], the LPDNet [13], the RIM [7], the RIM built with the IndRNN, the UNet [5], and conventional Compressed Sensing reconstruction [4]. The performance is evaluated on multi-modal MRI datasets applying different undersampling strategies. As a clinical application, we focused on reconstructing (out-of-training distribution) FLAIR data of Multiple Sclerosis patients. Finally, reconstruction times are also assessed as a critical aspect of improving clinical workflow.

## 2. Methods

In this section, first in 2.1, the MRI acquisition process is introduced. In 2.2, the background on solving the inverse problem of accelerated MRI reconstruction through a Bayesian approach is set. In 2.2.1 and 2.2.2, unrolled optimization by gradient descent is reviewed via the Recurrent Inference Machines (RIM) and the End-to-End Variational Network (E2EVN). The Cascades of Independently Recurrent Inference Machines (CIRIM) is then proposed in 2.2.3, to expand further de-aliasing capabilities of a deep trainable RNN. Furthermore, assessment of Data Consistency (DC) is performed in 2.2.2 and 2.2.3 to evaluate to what extent the performance of networks depends on the cascades or the DC formulation, or both. In 2.2.4, the loss function is explained with respect to the network's architecture. In 2.3, the experiments are described, i.e., the used datasets, the computed evaluation metrics, and the hyperparameters to be optimized.

## 2.1. Accelerated MRI Acquisition

The process of accelerating the MRI acquisition can be described through a forward model. Let the true image be denoted by $x \in \mathbb{C}^n$, with $n = n_x \times n_y$, and let $y \in \mathbb{C}^m$, with $m << n$, be the set of the sparsely sampled data in k-space. The forward model describes how the measured data are obtained from an underlying reference image. For the $i$-th coil of $c$ receiver coils, the forward model is formulated as:

$$y_i = \mathcal{A}(x) + \sigma_i, i = 1, \dots, c, \tag{1}$$

in which $\mathcal{A}: \mathbb{C}^n \mapsto \mathbb{C}^{n \times n_c}$ denotes the linear forward operator modeling the sub-sampling process of multicoil data and $\sigma_i \in \mathbb{C}^n$ denotes the measured noise for the $i$-th coil. $\mathcal{A}$ is given by

$$\mathcal{A} = \mathcal{P} \circ \mathcal{F} \circ \varepsilon. \tag{2}$$

Here, $\mathcal{P}$ is a sub-sampling mask selecting a fraction of samples to reduce scan time. $\mathcal{F}$ denotes the Fourier transform, projecting the image onto k-space. $\varepsilon: \mathbb{C}^n \times \mathbb{C}^{n \times n_c} \mapsto \mathbb{C}^{n \times n_c}$ denotes the expand operator, transforming $x$ into $x_c$ multicoil images and is given by

$$\varepsilon(x) = (\mathcal{S}_0 \circ x, \dots, \mathcal{S}_c \circ x) = (x_0, \dots, x_c), \tag{3}$$

where $\mathcal{S}_i$ are the coil sensitivity maps, a diagonal matrix representing the spatial sensitivities that scale every pixel of the reference image by a complex number.

The adjoint backward operator of $\mathcal{A}$ in (2), projecting $y$ onto image space, is given by

$$\mathcal{A}^* = \rho \circ \mathcal{F}^{-1} \circ \mathcal{P}^T, \tag{4}$$

where $\mathcal{F}^{-1}$ denotes the inverse Fourier transform, and $\rho: \mathbb{C}^{n \times n_c} \times \mathbb{C}^{n \times n_c} \mapsto \mathbb{C}^n$ denotes the reduce operator that serves for combining the multicoil images $x_c$ into $x$. $\rho$ is given by

$$\rho(x_0, \dots, x_c) = \sum_{i=1}^{c} \mathcal{S}_i^{\mathcal{H}} \circ x_i, \tag{5}$$

with $\mathcal{H}$ representing Hermitian complex conjugation.

## 2.2. The Inverse Problem of Accelerated MRI Reconstruction

The objective when solving the inverse problem of accelerated MRI reconstruction (Figure 1) is to map the sparsely sampled k-space measurements to an unaliased, highly accurate image.

The inverse transformation of restoring the true image from the set of the sparsely sampled measurements can be found through the Maximum A Posteriori (MAP) estimator, given by

$$x_{\text{MAP}} = \text{argmax}_x\{\log p(y|x) + \log p(x)\}, \quad (6)$$

which is the maximization of the sum of the log-likelihood and the log-prior distribution of $y$ and $x$, respectively. The log-likelihood expresses the log probability that k-space data $y$ are obtained given an image $x$, yielding a data fidelity term derived from the posterior $p(y|x)$. The log-prior distribution regularizes the solution by representing an MR-image's most likely appearance.

Conventionally, Eq. (6) is reformulated as the following optimization problem,

$$x_{\text{MAP}} = \text{argmin}_x\left\{\sum_{i=1}^{c} d(y_i, \mathcal{A}(x)) + \lambda \mathcal{R}(x)\right\}, \quad (7)$$

where $d$ ensures data consistency between the reconstruction and the measurements, reflecting the error distribution given by the log-likelihood distribution in Eq. (6). $\mathcal{R}$ is a regularizer weighted by λ, which constrains the solution space by incorporating prior knowledge over $x$.

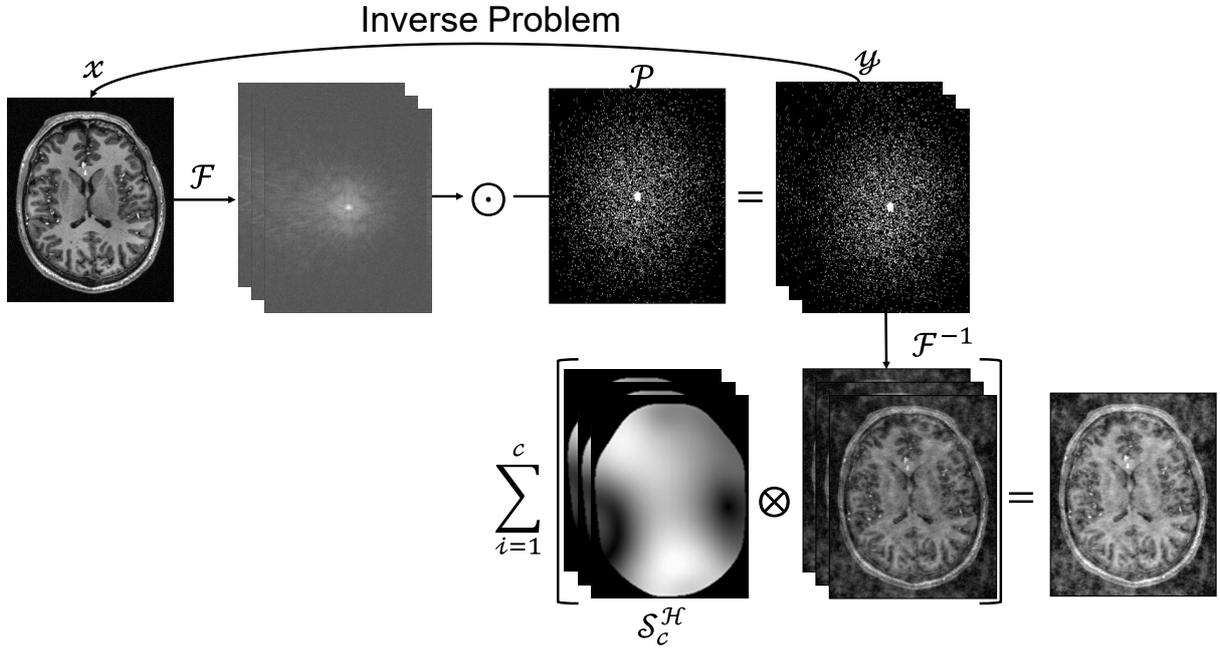

Figure 1: The objective in accelerated MRI reconstruction is to solve the inverse problem of recovering an unaliased image ($x$) from a set of sparsely sampled measurements ($y$). A forward model starts from the true image representation ($x$) (top-first), measured over multiple receiver coils ($S$) (bottom-first image). It is Fourier transformed to k-space (top-second) and sub-sampled using a mask ($\mathcal{P}$) (top-third) to obtain sparsely sampled measurements ($y$) (top-fourth). Through the inverse Fourier transform (bottom-second) and after combining with coil sensitivity maps (bottom-first), an aliased image is obtained (bottom-third).

Assuming Gaussian distributed data and ignoring the regularization term in Eq. (7), the negative log-likelihood is:

$$\log p(y|x) = \frac{1}{\sigma^2} \sum_{i=1}^{c} \|\mathcal{A}(x) - y_i\|_2^2. \qquad (8)$$

### 2.2.1. Recurrent Inference Machines (RIM)

The Recurrent Inference Machines (RIM) [7] were originally proposed as a general inverse problem solver. The RIM targets iterative optimization of a model with a complex-valued parametrization, requiring taking derivatives with respect to a complex variable. This can be achieved using the Wirtinger- or $\mathbb{CR}$- calculus [48–50]. Gradient descent is performed by us using the Wirtinger derivative, to yield a real-valued cost function of complex values. The unrolled scheme for generating updates is presented in Figure 2.

Non-convex optimization can be performed based on the approach by [51]. The update rules are learned by the optimizer $h$, which has its own set of parameters $\phi$. Formulating Eq. (6) accordingly, resulting updates are of the form

$$x_{\tau+1} = x_\tau + h_\phi(\nabla_{y|x_\tau}, x_\tau), \qquad (9)$$

at iteration $\tau$ and for a (a priori set) total number of iterations $\mathcal{T}$.

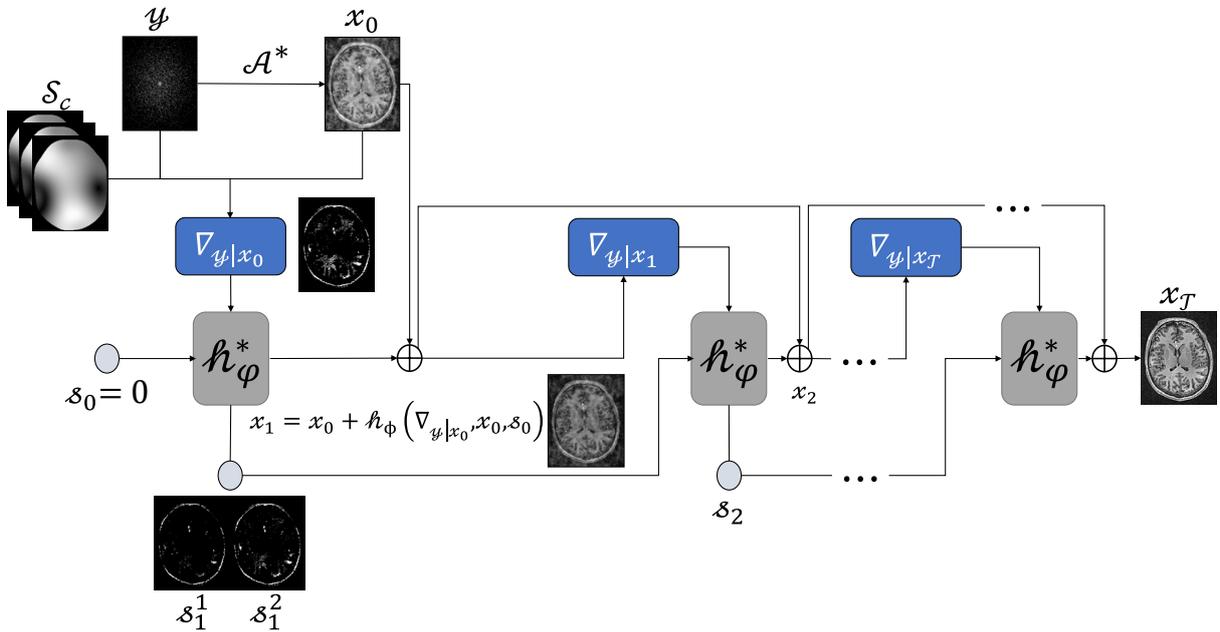

Figure 2: The Recurrent Inference Machines unrolled over two iterations. The inputs to the model are the set of sparsely sampled measurements ($y$) (top, second image), the coil sensitivities maps ($S_c$) (top, first image), and the initial estimation ($x_0$) (top, third image) for the estimation of the log-likelihood gradient (llg) ($\nabla_{y|x_0}$). The llg is passed through a network to produce updates; the network maintains hidden states initialized by $s_0 = 0$, $s_1 = 0$. At each iteration ($\tau$) the network updates itself and after total ($\mathcal{T}$) iterations produces the final estimation ($x_\mathcal{T}$) (rightmost).

The gradient of the log-likelihood function is given by

$$\nabla_{y|x} := \frac{1}{\sigma^2} \mathcal{A}^*(\mathcal{A}(x) - y). \qquad (10)$$

The advantage of the RIM is the explicit modeling of the update rule $h_\phi$ using a Recurrent Neural Network (RNN). In addition to the gradient information, the model is aware of the position of the estimation in variable space Eq. (9).

By inserting Eq. (10) into Eq. (9), the update equations are obtained, given by

$$s_0 = 0, \qquad x_0 = \mathcal{A}^*(y),$$
$$s_{\tau+1} = h^*_\phi(\nabla_{y|x_\tau}, x_\tau, s_\tau), \qquad x_{\tau+1} = x_\tau + h_\phi(\nabla_{y|x_\tau}, x_\tau, s_{\tau+1}). \qquad (11)$$

where $h^*_\phi$ is the updated model for state variable $s$. Eq. (11) reflects that not the prior is explicitly evaluated, but instead its gradient when performing updates. The step size is learned implicitly in combination with the prior. Therefore, $h_\phi$ also acts as regularizer $\mathcal{R}$ in Eq. (7). Observe that the RIM contains latent (hidden) states, representing the recurrent aspects of the network

### 2.2.2. End-to-End Variational Network (E2EVN)

The Variational Network (VN) [6] introduces a mapping to real-valued numbers, going from mapping $\mathbb{C}^n \mapsto \mathbb{C}^m$ to mapping $\mathbb{R}^{2n} \mapsto \mathbb{R}^{2m}$. $x$ can be computed by least-squares minimization in Eq. (8). As originally proposed in [52] and adapted by the VN, the idea is to perform gradient descent through the iterative Landweber algorithm. By defining a regularizer $\mathcal{R}$, Eq. (7) can be formulated as a trainable gradient scheme with time-varying parameters.

The End-to-End Variational Network (E2EVN) [21] uses a UNet as regularizer ($\mathcal{R}_{\text{UNet}}$), whose parameters are learned from the data. Unrolled optimization of the regularized problem in Eq. (7) is performed through cascades, given by

$$\hat{x}_{k+1} = \lambda \mathcal{R}_{\text{UNet}_k}\left(\mathcal{A}^*(y_k)\right), \qquad (12)$$

for cascade $k$, with $1 \leq k \leq \mathcal{K}$ for a total number of $\mathcal{K}$ cascades. Next, an explicitly formulated data consistency step applies k-space corrections. This step is given by

$$y_{k+1} = y_k - d(y_k - y) - \mathcal{A}(\hat{x}_{k+1}), \qquad (13)$$

where $d(y_k - y)$ is a soft DC term, with a weighting factor $d$. The optimization is initialized with the (sparsely sampled) measurement data, $y_{k=1} = y$. The eventual image is obtained via the adjoint operator $x_\mathcal{K} = \mathcal{A}^*(y_\mathcal{K})$.

In this paper, we test omitting the DC step, in Eq. (13), and evaluate if the network's performance is more dependent on the cascades or the gradient step. In that case, updates are given by Eq. (12). Note that the cascades effectively yield sequentially connected VN blocks, targeting de-aliasing (Figure 3).

The complex-valued image to complex-valued image mapping is performed in image space by concatenating the real and imaginary parts along the coil dimension. After the regularizer's update in Eq. (12), the image is reshaped to have the real and imaginary parts stacked to a complex (last) dimension.

### 2.2.3. Cascades of Independently Recurrent Inference Machines (CIRIM)

We now propose Cascades of Independently Recurrent Inference Machines (CIRIM), consisting of sequentially connected RIM blocks (Figure 3). The cascades allow building a deep RNN without vanishing or exploding gradients issues and further evaluate Eq. (9) through $\mathcal{K}$ cascades. As such the RIM acts as regularizer ($\mathcal{R}_{RIM}$), while the updates to the CIRIM are given by

$$\hat{x}_{k+1} = x_k + \lambda \mathcal{R}_{RIM_k}(x_k), \tag{14}$$

for cascade $k$, with $1 \leq k \leq \mathcal{K}$.

In previous work [7,9], the Gated Recurrent Unit (GRU) [53] was used as recurrent unit for the RIM. A key novelty of our approach is to include the Independently Recurrent Neural Networks (IndRNN) [47] as a more efficient unit for balancing the network's complexity while increasing the number of trainable parameters through the cascades.

Through the cascades the network's size has increased, but it is unclear whether either implicitly evaluating data consistency through the log-likelihood gradient in Eq. (10) is adequate, or an additional learned gradient step is needed to constrain the solution space further. In a similar manner as in Eq. (13), we assess enforcing DC explicitly and interleaved between the cascades. By doing so, we aim to understand to what extent the network's performance and de-aliasing capabilities depend on the cascades or the formulation of the DC.

Then, the updated prediction of the model is given by

$$\hat{x}_{k+1} = \mathcal{A}^*\big(y_k - d(y_k - y) - \mathcal{A}(\hat{x}_{k+1})\big), \tag{15}$$

with $x_{\mathcal{K}} = \mathcal{A}^*(y_{\mathcal{K}})$. If this DC step is omitted, updates to the CIRIM are given by Eq. (14). Implementation notation for the recurrent units can be found in the Appendix.

### 2.2.4. Loss function

For calculating the loss, we compare magnitude images derived from the complex-valued estimations $\hat{x}$ against the fully sampled reference $x$. As a loss function, we choose the $\ell_1$-norm. The $\ell_1$-norm represents the sum of the absolute difference, given by

$$\mathcal{L}^{\ell_1}(\hat{x}) = |\hat{x} - x|. \tag{16}$$

For the E2EVN and other trained models, the loss is given by Eq. (16). For the CIRIM, the loss is weighted depending on the number of iterations and averaged over the $\mathcal{K}$ cascades, to emphasize the predictions of the later iterations. The loss is then formulated as

$$\mathcal{L}^{\ell_1}(\hat{x}) = \frac{\sum_{i=1}^{c}\left(\frac{1}{q\mathcal{T}}\sum_{\tau=1}^{\mathcal{T}} w_\tau |\hat{x}_\tau - x|\right)}{\mathcal{K}}, \tag{17}$$

where $q$ is the total number of pixels of the image and $w_\tau$ is a weighting vector of length $\mathcal{T}$ prioritizing the loss at later time-steps. The weights are calculated by setting $w_\tau = 10^{-\frac{\mathcal{T}-\tau}{\mathcal{T}-1}}$.

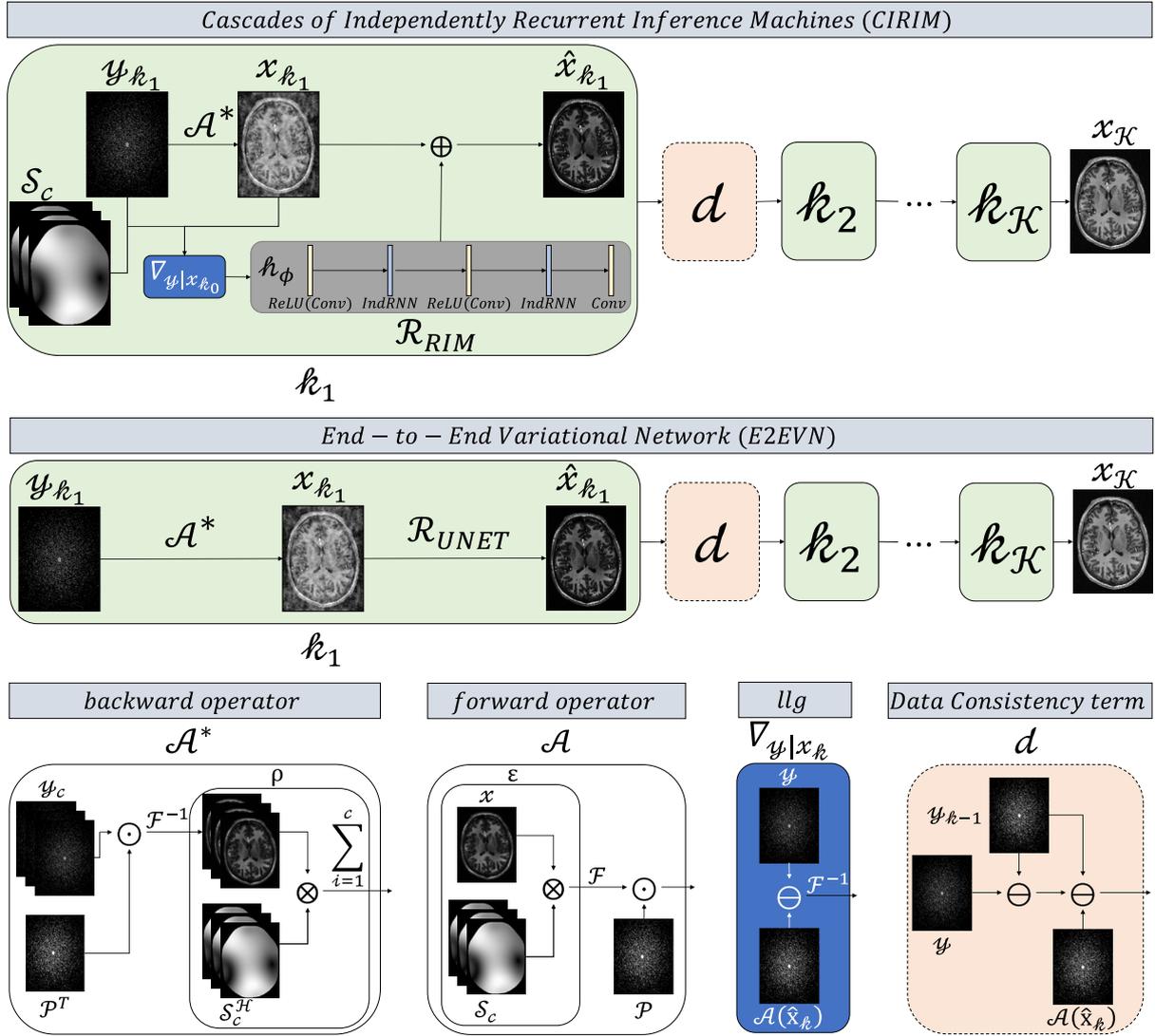

Figure 3: Overview scheme for performing unrolled optimization through cascades. The first row represents the Cascades of Independently Recurrent Inference Machines (**CIRIM**), in which a RIM is used as a regularizer ($\mathcal{R}_{RIM}$). The prediction ($\hat{x}_{k_1}$) of the first cascade ($k_1$) is given as input to the subsequent cascade ($k_2$), while an (optional) additional data consistency step can be performed through an explicitly formulated term ($d$). After ($\mathcal{K}$) cascades the network returns the final prediction ($x_{\mathcal{K}}$). The second row depicts the End-to-end Variational Network (**E2EVN**), where a UNet is used as a regularizer ($\mathcal{R}_{UNET}$). Similarly, as for the CIRIM, the updates are passed through the cascades and the data consistency step. In the third row, first, the backward operator ($\mathcal{A}^*$) is shown, transforming multicoil k-space measurements onto a coil-combined image; second, the forward operator ($\mathcal{A}$) is depicted, transforming a coil-combined image into multicoil k-space measurements; third, the log-likelihood gradient ($\nabla_{y|x_k}$) reflects the implicit gradient step of the RIM and fourth, the (optional) interleaved between the cascades DC term ($d$) is presented, enforcing an explicit gradient step to the CIRIM and the E2EVN.

## 2.3. Experiments

For our experiments, we used multiple datasets as described in 2.3.1. Scanning parameters of these datasets are given in Table 1.

Our experiments focused on assessment of the following aspects:

- A. Training and validation in fully sampled and retrospectively undersampled data. The undersampling strategy is described in 2.3.2.

B. Independent evaluation in prospectively undersampled data of Multiple Sclerosis patients containing white matter lesions.

We trained and compared the CIRIM and the E2VN to the LPDNet, the KIKINet, and the CascadeNet [13,15,26], the hyperparameters of which are described in 2.3.3. For comparing the performance of the methods regarding assessment (A) we chose the Structural Similarity Index Measure (SSIM) [54] and the Peak Signal-to-Noise Ratio (PSNR). For assessment (B), we calculated the Contrast Resolution (CR), the noise in the White Matter (WMN), the noise in the background (BGN), and a resulted Weighted Average (WA). The metrics are described in 2.3.4.

## 2.3.1. Datasets

For assessment (A), three fully sampled raw complex-valued multi-coil datasets were obtained. The first dataset was acquired in-house. To this end, eleven healthy subjects were included, from whom written informed consent (under an institutionally approved protocol) was obtained beforehand. The ethics board of Amsterdam UMC declared that this study was exempt from IRB approval. All eleven subjects were scanned by performing 3D $T_1$-weighted brain imaging on a 3.0T Philips Ingenia scanner (Philips Healthcare, Best, The Netherlands) in Amsterdam UMC. The data were visually checked to ascertain that they were not affected by motion artifacts. After scanning, raw data were exported and stored for offline reconstruction experiments. The training set was composed of ten subjects (approximately 3,000 slices) and the validation set of one subject (approximately 300 slices).

The second dataset consisted of 451 2D multislice FLAIR scans, publicly available through the fastMRI brains dataset [55]. The training set consisted of 344 scans (approximately 5,000 slices) and the validation set of 107 scans (approximately 500 slices). The number of coils varied from 2 to 24. The data were cropped in the image domain to 320 for the readout direction by the size of the phase encoding direction (varied from 213 to 320). The cropped images were visually evaluated to not crop any tissue (only air).

The third dataset was composed of 3D knee scans of 20 subjects, available on a public repository [56]. From these data, two subjects were discarded due to observed motion artifacts. The training set consisted of 17 subjects (approximately 12,000 slices) and the validation set of one subject (approximately 700 slices).

For all datasets, coil sensitivities were estimated using an autocalibration procedure called ecalib from the BART toolbox [57], which leverages the ESPIRiT algorithm [58]. For training and validation, slices were randomly selected by setting a random seed to enable deterministic behavior for all methods and ensure reproducibility. Note that the validation set was only used to calculate the loss at the end of each epoch and not included into the training set. Finally, all volumes were normalized to the maximum magnitude.

For assessment (B), testing the methods' ability to reconstruct unseen pathology, a dataset of 3D FLAIR data of Multiple Sclerosis patients with known white matter lesions was obtained. Data were prospectively undersampled with a factor of 7.5x based on a Variable-Density Poisson disk distribution. Originally these data were acquired on a 3.0T Philips Ingenia scanner (Philips Healthcare, Best, The Netherlands) in Amsterdam UMC, within the scope of a larger, ongoing study. The local ethics review board approved this study and patients provided informed consent prior to imaging. A fully-sampled reference scan was also acquired and used to estimate coil sensitivity maps using the caldir method of the BART toolbox [57]. The data were visually checked after which all subjects with motion artifacts were discarded, ending up including 18 patients (approximately 4000 slices).

Table 1: Scan parameters of each dataset used for different experiments. Target anatomy, contrast, scan, and field strength are given, with resolution (res), Field-of-View (FOV), time in minutes (with acceleration factor), number of coils (ncoils) and other scan parameters.

| scan / sequence | field strength | res (mm) | FOV (mm) | time (acc) | ncoils | parameters |
|---|---|---|---|---|---|---|
| Training, Validation | | | | | | |
| $T_1$-Brain / $T_1$ 3D MPRAGE | 3T | 1.0x1.0x1.0 | 256x256x240 | 10.8 (1x) | 32 | FA 9º, TFE-factor 150, TI=900ms |
| $T_2$-Knee / $T_2$ TSE | 3T | 0.5x0.5x0.6 | 160x160x154 | 15.3 (1x) | 8 | FA 90º, TR=1550ms, TE=25ms |
| FLAIR-Brain / 2D FLAIR | 1.5T / 3T | 0.7x0.7x5 | 220x220 | - (1x) | 2-24 | FA 150º, TR=9000ms, TE=78-126ms |
| Pathology study | | | | | | |
| MS FLAIR-Brain / 3D FLAIR | 3T | 1.0x1.0x1.1 | 224x224x190 | 4.5 (7.5x) | 32 | TR=4800ms, TE=350ms, TI=16500ms |

## 2.3.2. Undersampling

The masks for retrospective undersampling in assessment (A) were initially defined in 2D. As such the models trained on all modalities could also be used later for reconstructing high-resolution isotropic FLAIR data for assessment (B). The 3D datasets were first Fourier transformed along the frequency encoding axis and used as separate slices along the two-phase encoding axes. The 2D multislice FLAIR dataset was initially Fourier transformed along the frequency encoding axis and undersampled per slice in 2D, to train a model on an identical contrast as in assessment (B), while also having pathology present in the data.

All data were retrospectively undersampled in 2D by sampling k-space points from a Gaussian distribution with a Full Width at Half Maximum (FWHM) of 0.7, relative to the k-space dimensions. Hereby the sampling of low frequencies is prioritized whereas incoherent noise is created due to the random sampling. Note that in this way, we abide by the Compressed Sensing (CS) requirement of processing incoherently sampled data [3]. For autocalibration purposes, data points near the k-space center were fully sampled within an ellipse of which the half-axes were set to 2% of the fully sampled region. Acceleration factors of 4x, 6x, 8x, and 10x were used by randomly generating a sampling mask ($\mathcal{P}$) with according sampling density (both during training and validation).

To abide to the underlying sampling protocol, and to test the model's ability to reconstruct undersampled data in 1D, we performed an additional experiment with retrospective undersampling in just one dimension. Equidistant k-space points were sampled in the phase encoding direction [58]. The acceleration factor was set to four, while the central region was densely sampled retaining eight percent of the fully-sampled k-space.

### 2.3.3. Hyperparameters

For the CIRIM models, hyperparameters were selected as follows. The number of cascades $\mathcal{K}$ was set to 5, the number of channels to 64 for the recurrent and convolutional layers, and the number of iterations $\mathcal{T}$ to 8. The hyperparameter search for finding the optimal number of cascades is shown in the Supplementary Material. The kernel size of the first convolutional layer was set to $5 \times 5$ and to $3 \times 3$ for the second and third layers. The optimization of these hyperparameters is described elsewhere [7]. Next, we trained models on the $T_1$-Brain dataset, the $T_2$-Knee dataset, and the FLAIR-Brain dataset to realize the DC step from Eq. (15).

For the E2EVN models, we chose 8 cascades, 4 pooling layers, 18 channels for the convolutional layers, and included the DC step from Eq. (13). The hyperparameter search for finding the optimal number of cascades, pooling layers, and number of channels, is again shown in the Supplementary Material. Then, for further optimization, we experimented with training models on the $T_1$-Brain dataset, the $T_2$-Knee dataset, and the FLAIR-Brain dataset while omitting the DC step. The inputs to the UNet regularizer were padded for making the inputs square, setting the padding size to 11, and the outputs were unpadded for restoring the original input size.

For the baseline UNet, the number of input and output channels was set to 2. The number of channels for the convolutional layers was set to 64, and we chose 2 pooling layers. Similar to the E2EVN, the padding size was set to 11, while no dropout was applied. The selected hyperparameters for the UNet were motivated by the configuration in [59].

For the LPDNet, the KIKINet, and the CascadeNet, the choice of the hyperparameters was motivated from the baseline proposed models. For the LPDNet we used the same network architecture for both the primal and the dual part, being a UNet with 16 channels, 2 pooling layers, and padding size of 11, while no dropout was applied. The number of the primals, the duals, and the number of unrolled iterations was set to 5. Similarly, for the KIKINet, we used the UNet architecture for the k-space and the image space networks. The number of channels was set to 64, the number of pooling layers to 2, and the padding size to 11, without applying any dropout. Finally, for the CascadeNet the number of cascades set to 10, using a sequence of CNNs with 64 channels and depth size of 5, without applying batch normalization.

For all models, we applied the ADAM optimizer [60], setting the learning rate to 1e-3, except for the CascadeNet where the learning rate was set to 1e-5. The batch size was set to 1, allowing training on various input sizes. The data type was set to complex64 for complex-valued data and to float16 for real-valued data. For training models with 2D undersampling, the loss function for the CIRIM is given by Eq. (17) and for all models by Eq. (16). For training models with 1D undersampling, we used the SSIM as loss function, motivated by [19], as a better option for resolving artifacts introduced by equidistant undersampling.

CS reconstructions were performed using the BART toolbox [57]. Here we used Parallel-Imaging Compressed Sensing (PICS) with a $\ell_1$-wavelet sparsity transform. The regularization parameter was set to $\alpha = 0.005$, which was heuristically determined as a trade-off between aliasing noise and blurring. The maximum number of iterations was set to 60. We tested the reconstruction times of CS on the GPU (turning the -g flag on).

All experiments were performed on an Nvidia Tesla V100 with 32GB of memory. The code was implemented in PyTorch 1.9 [61] and PyTorch-Lighting 1.5.5 [62], on top of novel frameworks [63,64], and can be found at https://github.com/wdika/mridc.

### 2.3.4. Evaluation metrics

For quantitative evaluation of the fully-sampled measurements, we compared normalized magnitude images derived from the complex-valued estimations $x_\tau$ against the reference $x$ and calculated SSIM and PSNR metrics.

For evaluating robustness on the 3D FLAIR MS data, we computed the Contrast Resolution (CR), the noise in the White Matter (WMN), the noise in the background (BGN), and a resulted Weighted Average (WA) of those three metrics.

Since the data are not fully-sampled, the CR is an efficient metric to evaluate the signal level between the white matter and the lesions. To compute CR, lesion segmentations were performed using a pretrained Multi-View Convolutional Neural Network (MV-CNN). The MV-CNN was previously trained on combined Fast Imaging Employing Steady-state Acquisition (FIESTA), $T_2$-weighted and contrast-enhanced $T_1$-weighted data, for eye and tumor segmentation of retinoblastoma patients [65]. For the segmentation of the white matter, the Statistical Parametric Mapping (SPM) toolbox was used [66]. The mean lesion intensity was compared to that of presumed homogeneous surrounding white matter. To that end, the lesion masks were dilated by four voxels and intersected with the whole brain white matter mask. The CR is then defined as the difference between the lesion signal and the signal in the surrounding white matter, divided by the summation of them, given by

$$\text{CR} = \frac{s_{\text{lesion}} - s_{\text{WMSurroundingLesion}}}{s_{\text{lesion}} + s_{\text{WMSurroundingLesion}}}. \qquad (18)$$

The WMN is defined as the mode of the gradient magnitude image $x$, given by

$$\text{WMN} = \text{mode}(\nabla |x / \overline{x_{\text{WM}}}|), \qquad (19)$$

where $\overline{x_{\text{WM}}}$ is the mean WM intensity. The background noise (BGN) is computed as the 99-percentile value in the background region, being the complement of a tissue mask.

A Weighted Average (WA) was eventually defined as the combination of the CR, the WMN, and the BGN after scaling them to maximum value.

Finally, for every scan, the Signal-to-Noise Ratio (SNR) was calculated as follows,

$$\text{SNR} = \frac{\overline{t|x|}}{|\widetilde{y}|}, \qquad (20)$$

where $\overline{t|x|}$ is the mean value after thresholding the magnitude image $x$ to discard the background, and $|\widetilde{y}|$ the median magnitude value within a square region in the periphery of k-space, which was assumed to be dominated by imaging noise. The threshold $t$ was set using Otsu's method [67].

# 3. Results

Figure 4 shows SSIM and PSNR scores upon assessing Data Consistency (DC) explicitly and implicitly for the Cascades of Independently Recurrent Inference Machines (CIRIM) (Figure 4a) and the End-to-End Variational Network (E2EVN) (Figure 4b). The models were trained on the $T_1$-Brain dataset, the $T_2$-Knee dataset, and the FLAIR-Brain dataset.

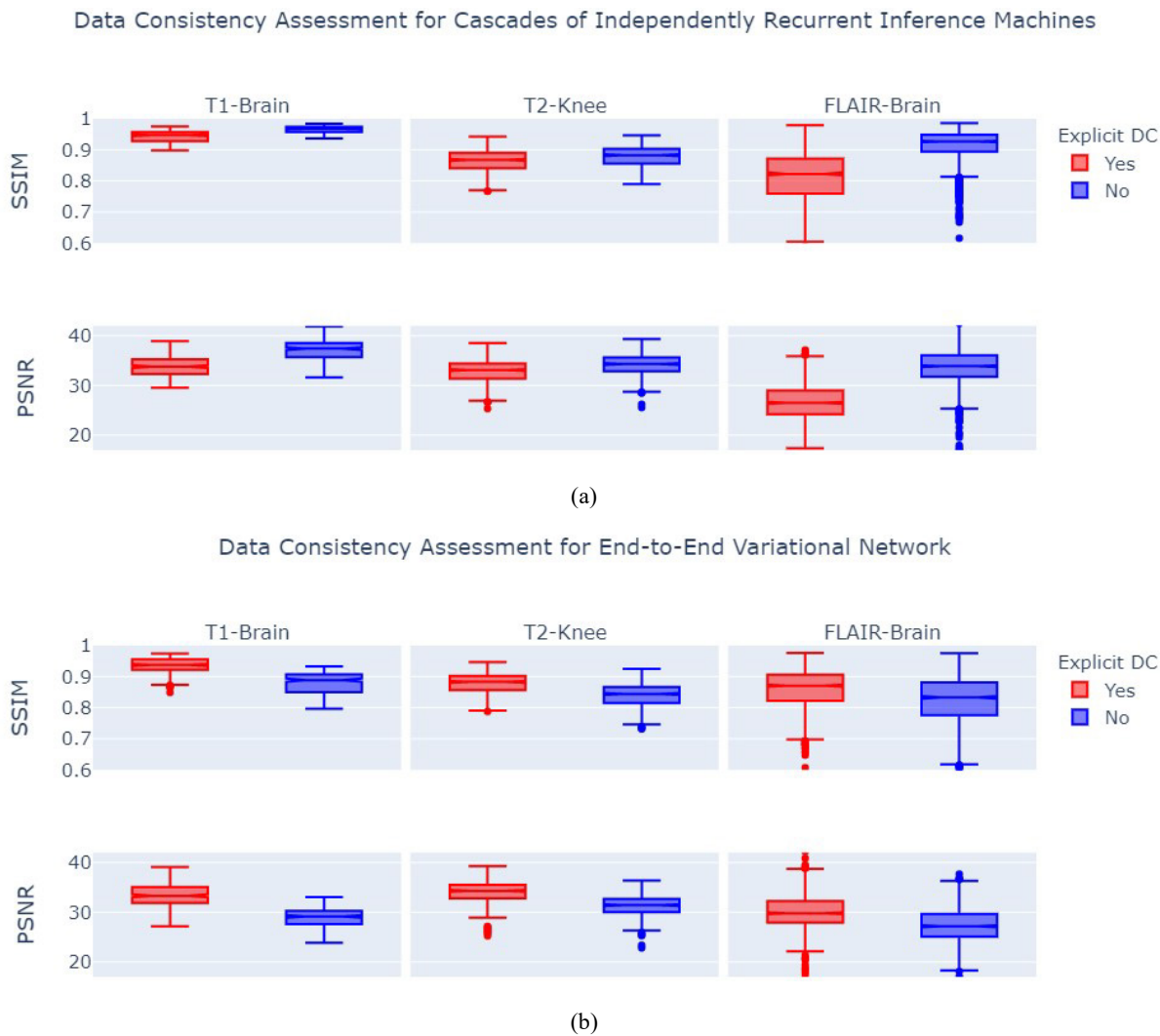

(a)

(b)

Figure 4: Data Consistency (DC) assessment for (a) Cascades of Independently Recurrent Inference Machines and (b) End-to-End Variational Network. DC is enforced both explicitly (red) and implicitly (blue). The first row represents SSIM scores and the second row PSNR scores. Performance is reported for models trained on the $T_1$-Brain dataset (first column), the $T_2$-Knee dataset (second column), and the FLAIR-Brain dataset (third column).

A qualitative evaluation of the CIRIM's and the E2EVN's performance on the trained datasets, accelerated with ten-times Gaussian 2D undersampling, is presented in Figure 5. The CIRIM performed significantly better than the E2EVN on the $T_1$-Brain and the FLAIR-Brain dataset. On the FLAIR-Brain dataset, the E2EVN failed to accurately reconstruct the center of brain, as well as to resolve noise in the White Matter lesion. On the $T_2$-Knee dataset, the two models performed comparably in terms of SSIM, while the CIRIM showed a slight improvement in PSNR.

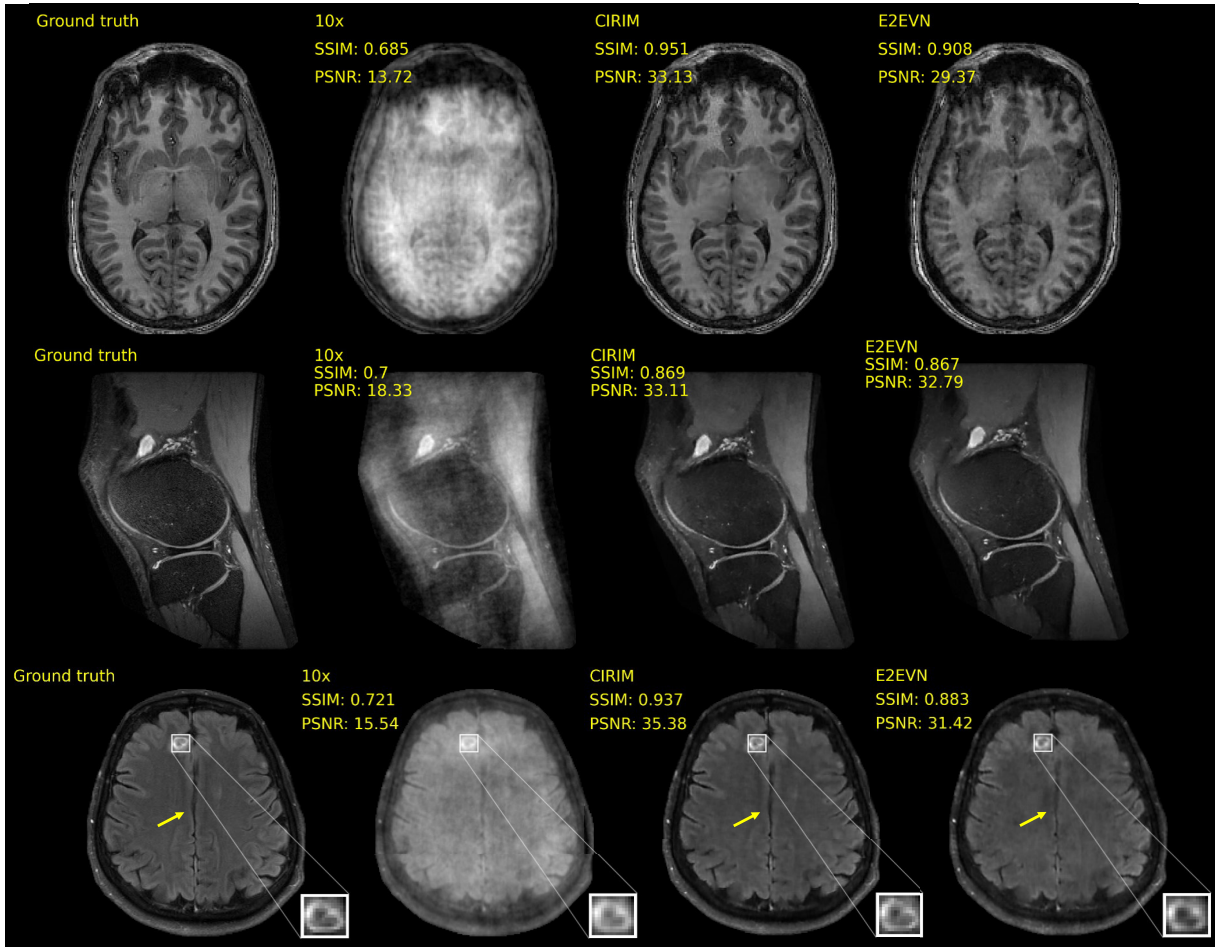

Figure 5: Comparison of the CIRIM (third column) to the E2EVN (fourth column) for reconstructing ten-times accelerated slices from the $T_1$-Brain dataset (first row, first and second image), the $T_2$-Knee dataset (second row, first and second image), and the FLAIR-Brain dataset (third row, first and second images). For the FLAIR-Brain dataset, the inset focuses on a reconstructed White Matter lesion; obtained through the fastMRI+ annotations (Zhao et al., 2021). The arrow points out to a region of interested.

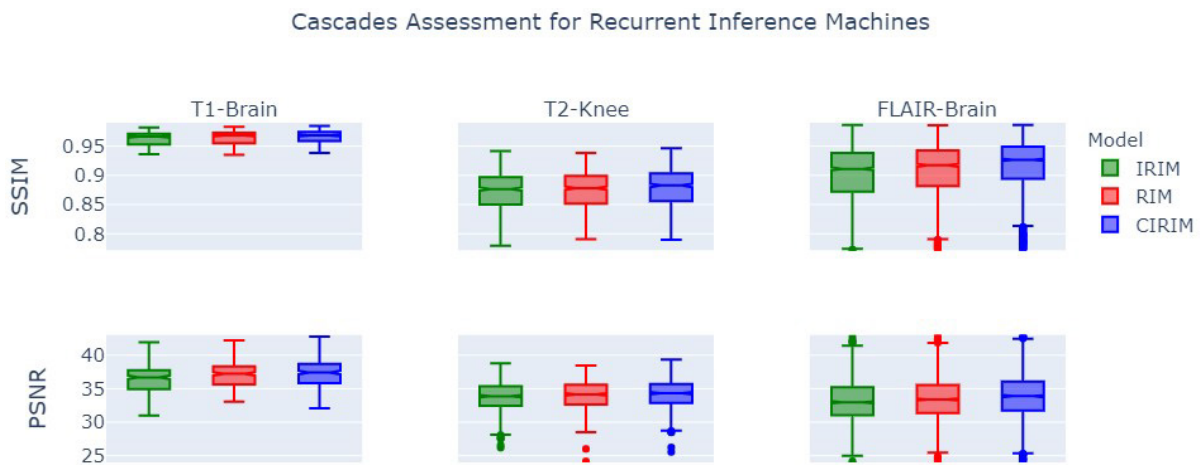

Figure 6: Comparison of the Cascades of Independently Recurrent Inference Machines (CIRIM) (blue color), to the Recurrent Inference Machines (RIM) (red color), and the Independently Recurrent Inference Machines (IRIM) (green color). Performance is reported for SSIM (first row) and PSNR (second row), on the $T_1$-Brain dataset (first column), the $T_2$-Knee dataset (second column), and the FLAIR-Brain dataset (third column).

In Figure 6, the CIRIM is compared to the RIM and the IRIM. SSIM and PSNR scores are reported for each model trained on the $T_1$-Brain dataset, the $T_2$-Knee dataset, and the FLAIR-Brain dataset. The IRIM performed slightly worse compared to the RIM, while the CIRIM performed best.

Table 2 collates overall performance of the methods on all training datasets ($T_1$-Brain, $T_2$-Knee, FLAIR-Brain). The methods were evaluated with ten-times accelerated data using Gaussian 2D masking, and four times accelerated equidistant 1D masking. For the FLAIR-Brain dataset we dropped the slices outside the head, containing no signal. The CIRIM performed best in all settings in terms of SSIM and PSNR, while only the E2EVN had comparable performance for the evaluation on the $T_2$-Knee dataset. Representative reconstructions can be found in the Supplementary Material, as well as further evaluation for four-, six-, and eight-times acceleration for Gaussian 2D undersampling.

Table 2: SSIM & PSNR scores of all methods evaluated on the $T_1$-Brain dataset (third and fourth column), the $T_2$-Knee dataset (fifth and sixth column), and the FLAIR-Brain dataset (seventh to tenth column). For all datasets performance is reported for ten times acceleration using Gaussian 2D undersampling. For the FLAIR-Brain dataset performance is also reported for four times acceleration using equidistant 1D undersampling (ninth and tenth column). The first column reports the method's name. The second column reports the total number of trainable parameters for each model. Best performing models are highlighted in bold. Methods are sorted in alphabetical order.

| Method | Params | $T_1$-Brain Gaussian 2D 10x | | $T_2$-Knee Gaussian 2D 10x | | FLAIR-Brain Gaussian 2D 10x | | FLAIR-Brain Equidistant 1D 4x | |
| --- | --- | --- | --- | --- | --- | --- | --- | --- | --- |
| | | SSIM ↑ | PSNR ↑ | SSIM ↑ | PSNR ↑ | SSIM ↑ | PSNR ↑ | SSIM ↑ | PSNR ↑ |
| CascadeNet | 1.1M | 0.922±0.042 | 30.2±1.5 | 0.859±0.045 | 32.2±2.6 | 0.872±0.106 | 29.9±5.0 | 0.913±0.038 | 30.3±4.6 |
| CIRIM | 264k | **0.966±0.015** | **35.8±0.5** | **0.877±0.039** | **33.7±2.3** | **0.906±0.101** | **32.8±6.2** | **0.942±0.065** | **34.3±3.2** |
| E2EVN | 19.6M | 0.940±0.023 | 31.8±1.4 | **0.877±0.039** | 33.5±2.5 | 0.855±0.108 | 29.1±4.8 | 0.930±0.062 | 31.3±5.1 |
| IRIM | 53k | 0.963±0.017 | 35.3±0.6 | 0.870±0.041 | 33.3±2.3 | 0.892±0.107 | 32.0±6.0 | 0.908±0.093 | 32.0±5.4 |
| KIKINet | 1.9M | 0.925±0.040 | 31.1±1.3 | 0.842±0.045 | 32.1±2.0 | 0.829±0.113 | 28.4±4.8 | 0.919±0.065 | 30.5±4.6 |
| LPDNet | 118k | 0.960±0.016 | 35.0±0.4 | 0.873±0.038 | 33.5±2.0 | 0.858±0.011 | 29.7±4.7 | 0.938±0.061 | 32.3±5.3 |
| PICS | | 0.866±0.032 | 30.9±0.7 | 0.729±0.041 | 29.7±4.3 | 0.816±0.174 | 29.2±7.6 | 0.876±0.068 | 30.0±4.2 |
| RIM | 94k | 0.963±0.017 | 35.3±0.4 | 0.872±0.040 | 33.5±2.3 | 0.898±0.103 | 32.3±6.1 | 0.934±0.069 | 33.4±3.2 |
| UNet | 1.9M | 0.874±0.049 | 26.6±3.1 | 0.846±0.048 | 31.4±3.4 | 0.795±0.116 | 26.9±4.1 | 0.909±0.064 | 29.4±4.3 |
| Zero-Filled | | 0.766±0.084 | 17.3±2.0 | 0.674±0.031 | 17.3±1.1 | 0.703±0.120 | 16.8±4.2 | 0.806±0.062 | 21.5±3.8 |

The trained models on each dataset and undersampling scheme were used to evaluate performance on out-of-training distribution data, containing Multiple Sclerosis (MS) lesions. As summarized in Table 3, the performance is evaluated quantitatively by measuring the Contrast Resolution (CR) of the reconstructed lesions, the White Matter Noise (WMN), the background noise (BGN) and a Weighted Average (WA). A combination of high CR, low WMN and low BGN yields a low WA and reflects highly accurate reconstruction (Figure 6, Figure S7, Figure S8), such as in the CIRIM FLAIR-Brain model and PICS. The models trained on the FLAIR-Brain, the FLAIR-Brain 1D, and the $T_1$-Brain datasets scored high on CR and low on WMN compared to the $T_2$-Knee trained models. The CIRIM and the RIM achieved the lowest BGN. The CascadeNet, the E2EVN, the KIKINet, and the UNet models reported high BGN, in general corresponding to more aliased reconstruction. The LPDNet achieved high CR and relatively low WMN and BGN, but the observed reconstruction quality was poor. This also highlighted the need for combined metrics and qualitative evaluation to evaluate performance.

Table 3: Independent evaluation of model performance (first column) on the 3D FLAIR MS Brains dataset for different training datasets (second column). The reported figures collate: Contrast Resolution (CR, higher is better) of MS lesions, gradient magnitude White Matter Noise (WMN, lower is better), Background Noise (BGN, lower is better) and Weighted Average (WA, with negative CR and relative to maximum scores such that lower is better), respectively. For each model and dataset, the mean and standard deviation on each metric is given. The best scores are underlined and other high scores highlighted in bold. Methods are sorted in alphabetical order.

| FLAIR MS Brains – Variable Density Poisson 7.5x | | | | | |
|---|---|---|---|---|---|
| Method | Trained Dataset | CR ↑ | WMN ↓ | BGN ↓ | WA ↓ |
| CascadeNet | $T_1$-Brain | 0.128±0.028 | 0.135±0.022 | 0.292±0.078 | 1.08 |
| | $T_2$-Knee | 0.087±0.040 | 0.290±0.059 | 0.302±0.083 | 1.43 |
| | FLAIR-Brain | 0.145±0.030 | 0.126±0.016 | 0.265±0.071 | 0.96 |
| | FLAIR-Brain 1D | 0.139±0.025 | 0.121±0.016 | 0.309±0.068 | 1.05 |
| CIRIM | $T_1$-Brain | 0.179±0.025 | 0.145±0.030 | 0.172±0.092 | 0.69 |
| | $T_2$-Knee | 0.097±0.020 | 0.285±0.044 | 0.322±0.053 | 1.42 |
| | **FLAIR-Brain** | **0.183±0.025** | 0.131±0.029 | **<u>0.104±0.085</u>** | **<u>0.55</u>** |
| | FLAIR-Brain 1D | 0.173±0.030 | **0.110±0.017** | 0.137±0.074 | **0.62** |
| E2EVN | $T_1$-Brain | 0.145±0.034 | 0.144±0.010 | 0.359±0.095 | 1.13 |
| | $T_2$-Knee | 0.109±0.028 | 0.301±0.042 | 0.576±0.352 | 1.79 |
| | FLAIR-Brain | 0.159±0.041 | 0.116±0.014 | 0.358±0.064 | 1.03 |
| | FLAIR-Brain 1D | 0.134±0.035 | 0.141±0.020 | 0.360±0.058 | 1.17 |
| IRIM | $T_1$-Brain | 0.159±0.025 | 0.128±0.027 | 0.200±0.089 | 0.80 |
| | $T_2$-Knee | 0.078±0.021 | 0.260±0.122 | 0.348±0.061 | 1.51 |
| | FLAIR-Brain | 0.169±0.027 | 0.145±0.020 | 0.181±0.126 | 0.74 |
| | FLAIR-Brain 1D | 0.176±0.025 | 0.151±0.020 | 0.213±0.081 | 0.77 |
| KIKINet | $T_1$-Brain | 0.117±0.032 | 0.184±0.042 | 0.432±0.075 | 1.40 |
| | $T_2$-Knee | 0.149±0.026 | 0.235±0.032 | 0.294±0.087 | 1.10 |
| | FLAIR-Brain | 0.105±0.077 | 0.175±0.040 | 0.626±0.141 | 1.75 |
| | FLAIR-Brain 1D | 0.103±0.026 | 0.144±0.035 | 0.352±0.060 | 1.29 |
| LPDNet | $T_1$-Brain | <u>0.240±0.046</u> | 0.206±0.029 | 0.210±0.056 | **0.56** |
| | $T_2$-Knee | 0.030±0.151 | 0.126±0.031 | 0.204±0.065 | 1.34 |
| | FLAIR-Brain | 0.117±0.024 | **0.099±0.012** | 0.332±0.083 | 1.15 |
| | FLAIR-Brain 1D | 0.066±0.029 | 0.129±0.024 | 0.338±0.070 | 1.40 |
| RIM | $T_1$-Brain | 0.178±0.025 | 0.168±0.026 | 0.170±0.093 | 0.71 |
| | $T_2$-Knee | 0.091±0.036 | 0.149±0.030 | 0.251±0.091 | 1.18 |
| | FLAIR-Brain | **0.197±0.029** | 0.175±0.025 | **0.134±0.078** | **0.58** |
| | FLAIR-Brain 1D | **0.183±0.027** | 0.158±0.026 | 0.165±0.074 | 0.67 |
| UNet | $T_1$-Brain | 0.182±0.034 | 0.174±0.022 | 0.276±0.069 | 0.87 |
| | $T_2$-Knee | 0.125±0.040 | 0.924±0.084 | 0.285±0.089 | 1.93 |
| | FLAIR-Brain | 0.087±0.027 | <u>0.079±0.010</u> | 0.625±0.137 | 1.72 |
| | FLAIR-Brain 1D | 0.065±0.021 | 0.105±0.023 | 0.348 ±0.070 | 1.40 |
| PICS | | 0.178±0.025 | 0.140±0.018 | **0.147±0.092** | 0.64 |
| Zero-Filled | | 0.072±0.023 | 0.092±0.017 | 0.372±0.064 | 1.39 |

Figure 7 shows reconstructions of a coronal slice from the MS FLAIR-Brain dataset. Visually, the CIRIM, PICS, RIM, and IRIM reconstructions appear similar. The E2EVN and the CascadeNet showed inhomogeneous intensities and high contrast deviations. The LPDNet showed more aliased reconstructions, with lower contrast levels. The KIKINet and the UNet seemed in our experiments not able to resolve background noise and in general resulted in more distorted images. Example reconstructions of two more subjects including axial and sagittal plane reconstructions can be found in the Supplementary Material.

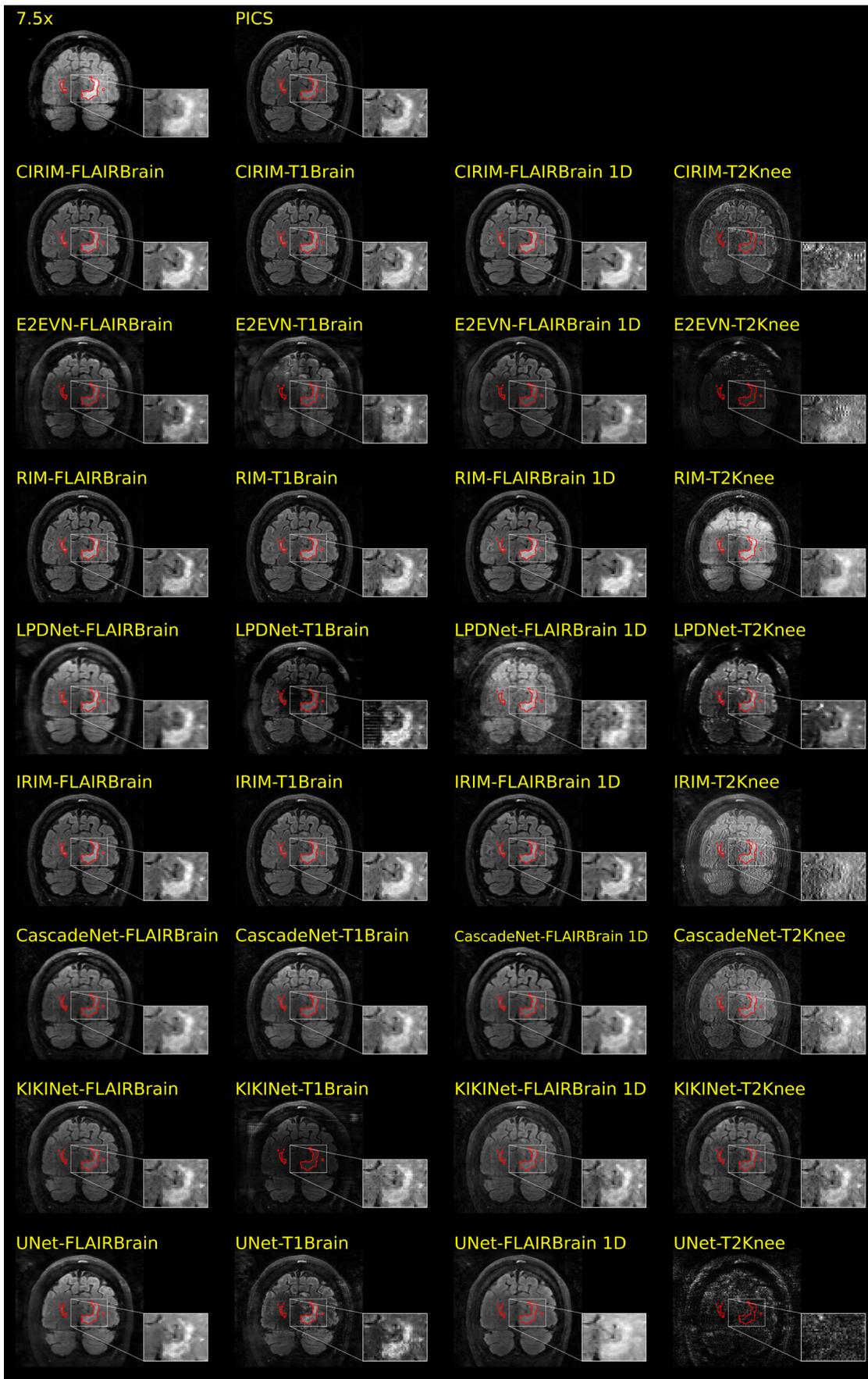

Figure 7: Reconstructions of a representative coronal slice of a 7.5x accelerated 3D FLAIR scan of a MS patient. Segmented MS lesions are depicted with red colored contours. Shown is the aliased linear reconstruction (first row-first image), PICS (first row-second image), and models' reconstructions on each trained scheme: the FLAIR-Brain dataset with Gaussian 2D undersampling (second-last row, first column), the $T_1$-Brain dataset with Gaussian 2D undersampling (second-last row, second column), the FLAIR-Brain dataset with equidistant 1D undersampling (second-last row, third column), and the $T_2$-Knee dataset with Gaussian 2D undersampling (second-last row, fourth column). The inset on the right bottom of each reconstruction focuses on a lesion region with high spatial detail.

Finally, in Figure 8, the reconstruction times of all methods are reported. As input, one volume from the trained fastMRI FLAIR brains dataset was taken, consisting of fifteen slices cropped to a matrix size of 320×320. The KIKINet, PICS, and the LPDNet were the slowest methods, requiring 247 ms, 245 ms, and 237 ms respectively to reconstruct the volume. The CIRIM needed 139 ms, the RIM 48 ms, the E2EVN 44 ms, the CascadeNet 42 ms, the IRIM 28 ms, and the UNet 8 ms.

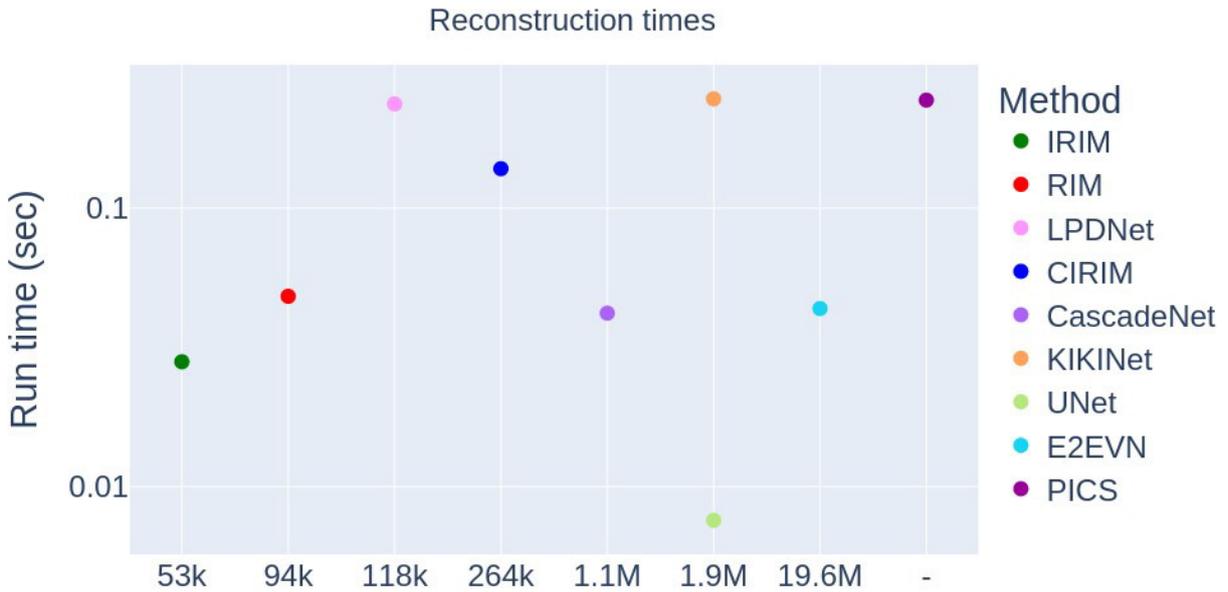

Figure 8: Inference times for reconstructing one volume from the FLAIR brains dataset using different methods. The x-axis represents methods' number of trainable parameters. The y-axis shows the run time in seconds.

## 4. Discussion

In this paper, we proposed the Cascades of Independently Recurrent Inference Machines (CIRIM), for a balanced increase in model complexity while maintaining generalization capabilities. We assessed Data Consistency (DC) both implicitly through unrolled optimization by gradient descent and explicitly by a formulated term. Robustness was evaluated by reconstructing sparsely sampled MRI data containing unseen pathology. The CIRIM was extensively compared to another unrolled network, the End-to-End Variational Network (E2EVN), and a range of other methods.

In experiments reconstructing brain and knee data containing different contrasts, the proposed CIRIM performed best, with promising generalization capabilities. On the $T_2$-knee dataset, the E2EVN performed equivalently to the CIRIM, while on the $T_1$-brain and the FLAIR-brain datasets for eight- and ten-times acceleration, the measured PSNR dropped by approximately 5% of what compared to what. Visually, this reflected in missing anatomical details such as vessels. The LPDNet, the RIM, and the IRIM performed comparable but slightly worse than the CIRIM. The CascadeNet and the KIKINet, dropped further in SSIM and PSNR on all trained datasets, resulting in more noisy reconstructions. PICS and the UNet showed most of the time overly smoothed results. Interestingly, for 1D undersampling the CascadeNet showed comparable performance to the CIRIM, but it was more sensitive to banding artifacts.

The RIM-based models (RIM, IRIM, CIRIM), trained on FLAIR and $T_1$-weighted brain data, and PICS, could accurately reconstruct Multiple Sclerosis lesions unseen during training. Spatial detail when reconstructing MS lesions was preserved with better denoised images, compared to, e.g., the CascadeNet. The E2EVN and the LPDNet did not show any significant improvement in this respect. The reason for such behavior might be that these scans, in contrast to the training data, came without a fully sampled center since a separate reference scan was acquired. This deviation from the training data could explain the lower performance of some of the models. Conditional deep priors tend to learn dealiasing of undersampled acquisitions on images that they have trained on. In such a situation, learning k-space corrections might be disadvantageous. The KIKINet and the UNet performed significantly worse than the other methods, thereby appearing to be sensitive to noisy inputs. Furthermore, the models trained on knees were inadequate in reconstructing MS lesions, indicating training anatomy preference rather than generalization. Remarkably, this was also realized by the performance of the networks trained on the FLAIR-Brain datasets with equidistant 1D undersampling. All models generalized well on reconstructing the MS data, despite the deviating undersampling scheme (variable density poisson sampling in 2D).

The SNR, the number of coils, and the size of the training dataset appeared to be important parameters that influenced performance. This is to be seen in the reported SSIM and PSNR scores. Here, the E2EVN models performed highest on the largest dataset, i.e., the $T_2$-weighted knee dataset, which contained approximately 12,000 slices. However, all models experienced lower performance due to lower SNR (17.1±4.5) and number of coils (8), compared to the $T_1$-weighted brain dataset (3,000 slices, SNR of 25.7±5.4, and 32 coils). The FLAIR brain dataset, despite its relatively high SNR (5,000 slices and SNR of 23.6±4.8), did not necessarily yield high quality in reconstructed images. The deviating number of coils (from 2 to 24), field strength (1.5T an 3T), and matrix sizes, resulted in a challenging dataset to converge with when training a model. In this situation, the advantage of implementing cascades was most apparent, making the CIRIM being robust with all tested acceleration factors (4x, 6x, 8x – Supplementary Material and 10x - Table 2). PICS and the UNet scored overall lower, illustrating that learning a prior with an efficient model is advantageous.

Importantly, our results show that the RIM-based models can reconstruct image details unseen during training. The RIM explicitly contains a formulation of the prior information of an MR image and acts as optimizer itself. Unrolled optimization is performed by gradient descent [9], such that DC is enforced implicitly. The CIRIM allows to further denoise the reconstructed images through the cascades without sharing parameters, similar to previously proposed deep cascading networks [26,68,69]. The cascades thereby allowed us to train an overall deep network of multiple connected RNNs that captures long-range dependencies while avoiding vanishing or exploding gradients. The E2EVN also performs unrolled optimization through cascades, but explicitly enforces DC with a formulated term.

Recent work has pointed out the importance of benchmarking and quantifying the performance of deep networks regarding the GPU memory required for training, the inference times, the applications, and the optimization [70–72]. With regard to inference times, methods such as the LPDNet and the KIKINet did not seem to improve in speed over the conventional CS algorithm, implemented on the GPU. The reason for these methods being slower is that they consist of deep feed-forward large convolutional layers. The RIM, the E2EVN, and the CascadeNet reduce reconstruction times by a factor of six compared to CS. Here, inference is performed over an iterative scheme, in which sharing of parameters is optimized either through time-steps or cascades. The IRIM and the UNet even further reduce the time by a factor of two and six, respectively. The CIRIM serves as a balanced deep network, being two times faster than the slowest methods and two times slower than the other cascading networks. The performance gain in further denoising and generalization capabilities may counterbalance the need for longer inference times.

## 5. Conclusion

The Cascades of Independently Recurrent Inference Machines (CIRIM) implicitly enforces Data Consistency (DC) when targeting unrolled optimization through gradient descent. The comparable E2EVN performed best when DC was explicitly enforced, performing well on the training distributions. However, it appeared to be inadequate on reconstructing out-of-training distribution data without a fully sampled center. The CIRIM performed best on all training datasets, tested undersampling schemes and acceleration factors. Also, it showed robust performance on reconstructing accelerated FLAIR data containing MS lesions, achieving good lesion contrast and efficient denoising compared to PICS, the baseline RIM and the IRIM. In contrast, methods such as the CascadeNet and the LPDNet were sensitive to highly noisy untrained data, showing limited generalization capabilities. The KIKINet and the UNet tended to oversimplify the reconstructed images, performing markedly worse than rest methods. To that extent, the impression is that evaluating the forward process of accelerated MRI reconstruction, frequently through time, is of great importance for generalization in other settings. The implemented cascades and the application of the RIM to a deeper network allowed backpropagation on a smaller number of time-steps but on higher frequency for each iteration. Thus, a key advantage of the CIRIM is that it maintains a very fair trade-off between reconstructed image quality and fast reconstruction times, which is crucial in the clinical workflow.

## Acknowledgments

This publication is based on the STAIRS project under the TKI-PPP program. The collaboration project is co-funded by the PPP Allowance made available by Health~Holland, Top Sector Life Sciences & Health, to stimulate public-private partnerships.

M.W.A. Caan is shareholder of Nico.Lab International Ltd.

# Appendix

## Gated Recurrent Unit (GRU)

The GRU has two gating units, the reset gate and the update gate. These gates control how the information flows in the network. The update gate regulates the update to a new hidden state, whereas the reset gate controls the information to forget. Both gates act in a probabilistic manner.

The activation of the reset gate $r$ at iteration $\tau$, for updating Eq. (9), is computed by

$$r_\tau = \sigma(\mathcal{W}_r[s_{\tau-1}, x_\tau] + b_r).$$

$\sigma$ is the logistic sigmoid function, $x_\tau$ and $s_{\tau-1}$ are the input and the previous hidden state, respectively. $\mathcal{W}_r$ and $b_r$ are the weights matrix and the learned bias vector.

Similarly, the update gate $z$ is computed by

$$z_\tau = \sigma(\mathcal{W}_z[s_{\tau-1}, x_\tau] + b_z).$$

The actual activation of the next hidden state $s_\tau$ is then computed by

$$s_\tau = (1 - z_\tau) \odot s_{\tau-1} + z_\tau \odot \tilde{s}_\tau,$$

where $\odot$ represents the Hadamard product and $\tilde{s}_\tau$ is given by

$$\tilde{s}_\tau = \tanh(\mathcal{W}_s[r_\tau \odot s_{\tau-1}, x_\tau] + b_s).$$

## Independently Recurrent Neural Network (IndRNN)

The IndRNN addresses gradient decay over iterations, following an independent neuron connectivity within a recurrent layer. The update on Eq. (9) and at iteration $\tau$ is given by

$$s_\tau = \sigma(\mathcal{W}_{x_\tau} + u \odot s_{\tau-1} + b),$$

where $\mathcal{W}$ is the weight for the current input, $u$ is the weight for the recurrent input, and $b$ is the bias vector.

# Supplementary Material

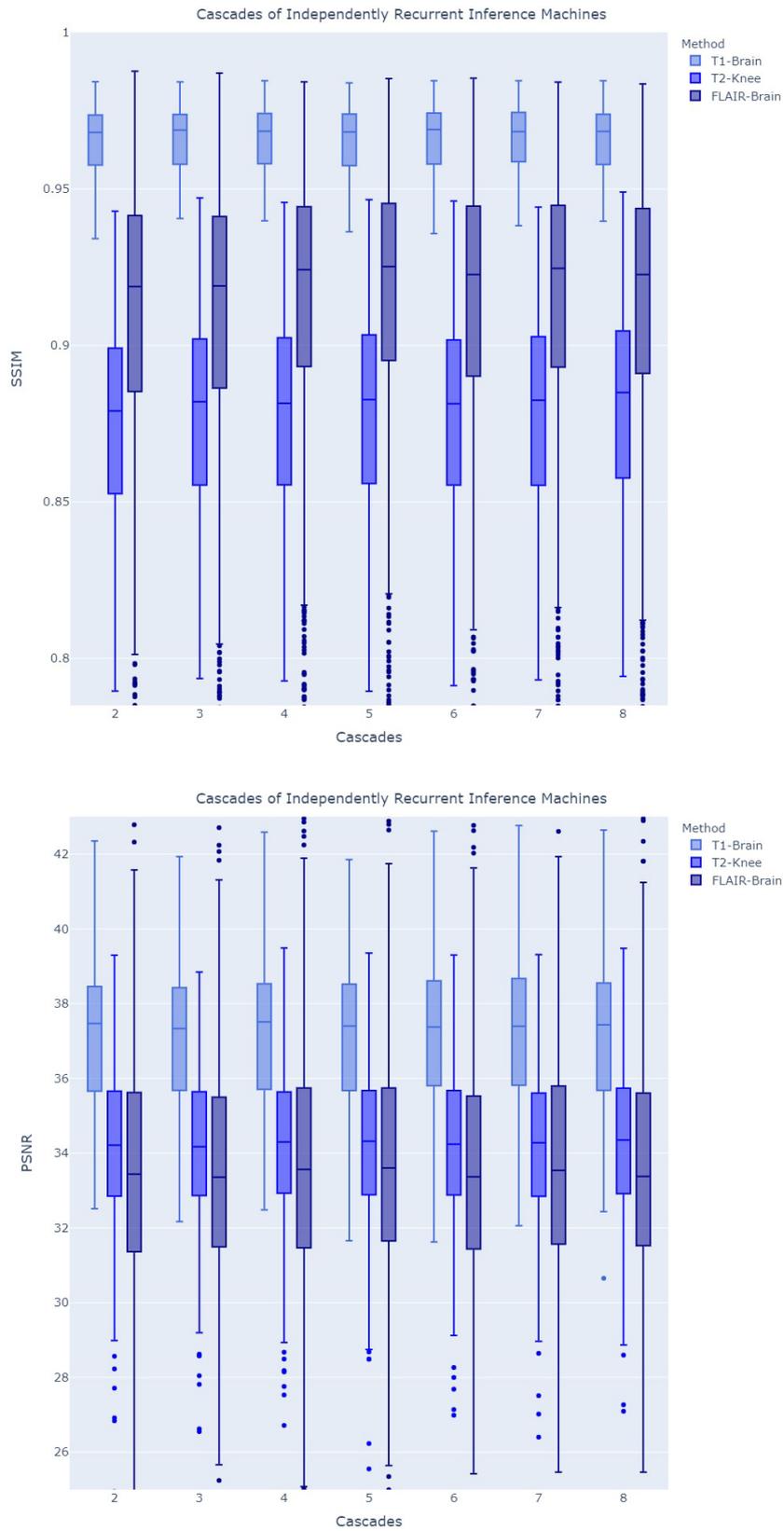

Figure S1: Comparison of varying number of cascades for the Cascades of Independently Recurrent Inference Machines, on the trained datasets ($T_1$-Brain, $T_2$-Knee, $FLAIR$-Brain) using Gaussian 2D 10x undersampling. Top figure reports SSIM scores and bottom figure PSNR scores.

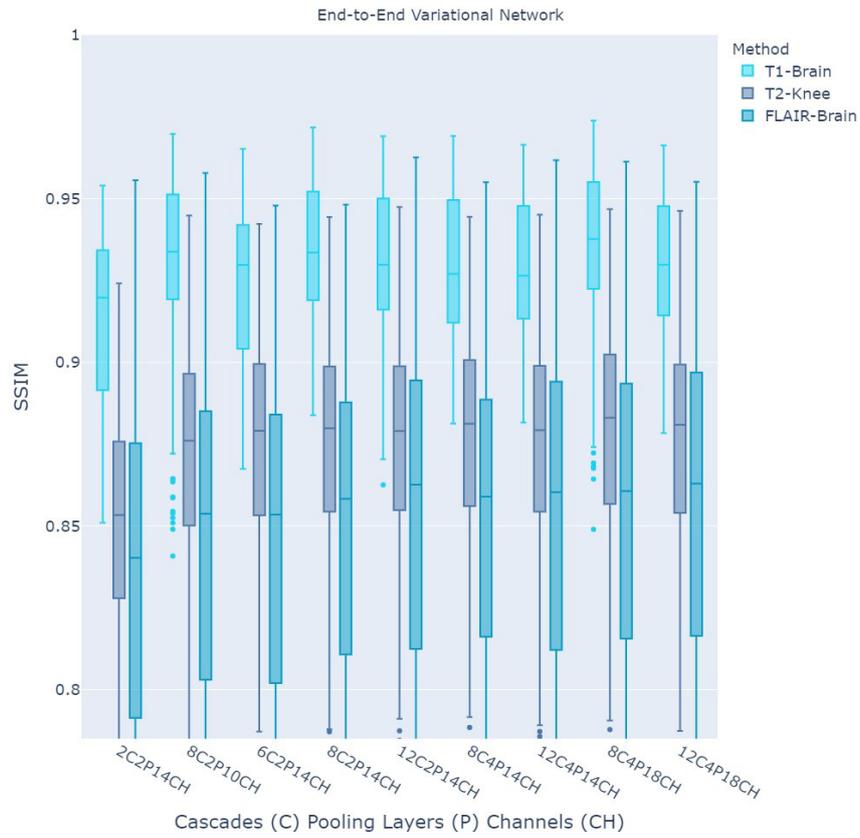

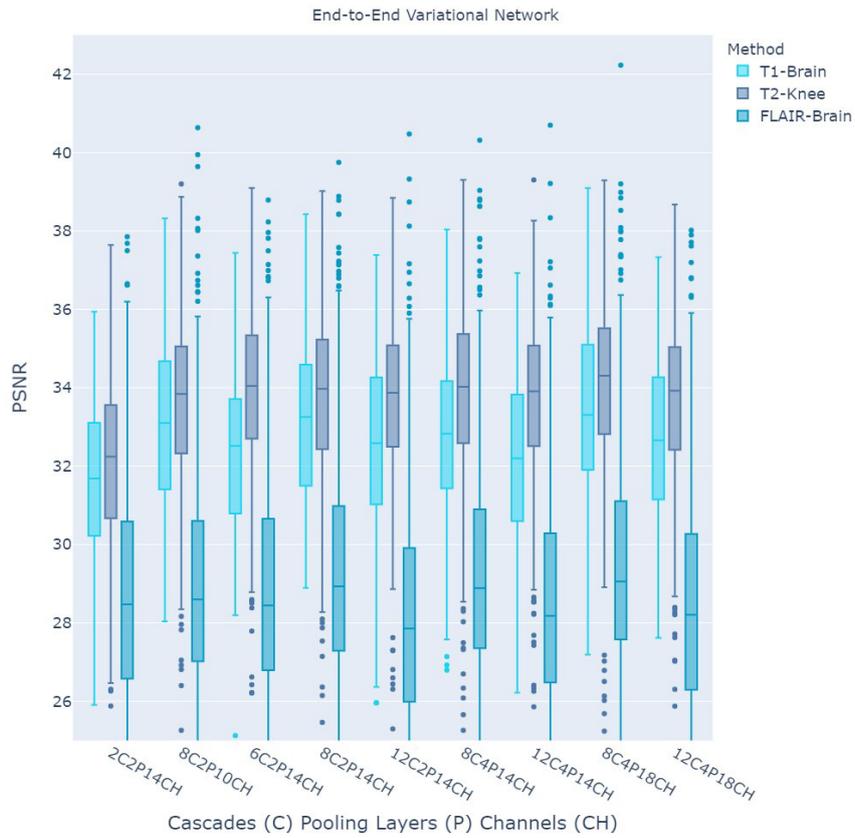

Figure S2: Comparison of varying number of cascades, pooling layers, and channels for the End-to-End Variational Network, on the trained datasets ($T_1$-Brain, $T_2$-Knee, $FLAIR$-Brain) using Gaussian 2D 10x undersampling. Top figure reports SSIM scores and bottom figure PSNR scores.

Table S1: SSIM & PSNR scores of all methods evaluated on the $T_1$-Brain dataset (third and fourth column), the $T_2$-Knee dataset (fifth and sixth column), and the FLAIR-Brain dataset (seventh, eighth, ninth and tenth column). For all datasets performance is reported for four times acceleration using Gaussian 2D undersampling. The second column reports the total number of trainable parameters for each model. Best performing models are highlighted in bold. Methods are sorted in alphabetical order.

| Method | Params | $T_1$-Brain Gaussian 2D 4x | | $T_2$-Knee Gaussian 2D 4x | | FLAIR-Brain Gaussian 2D 4x | |
|---|---|---|---|---|---|---|---|
| | | SSIM ↑ | PSNR ↑ | SSIM ↑ | PSNR ↑ | SSIM ↑ | PSNR ↑ |
| CascadeNet | 1.1M | 0.962±0.016 | 33.3±0.9 | 0.908±0.030 | 35.2±2.3 | 0.925±0.068 | 33.6±4.7 |
| CIRIM | 264k | **0.981±0.007** | **39.2±0.6** | **0.919±0.027** | 36.3±2.3 | **0.945±0.061** | **36.5±5.3** |
| E2EVN | 19.6M | 0.972±0.011 | 35.6±0.6 | **0.919±0.027** | **36.4±2.3** | 0.912±0.071 | 32.6±5.3 |
| IRIM | 53k | 0.980±0.008 | 38.9±0.6 | 0.912±0.029 | 35.8±2.2 | 0.937±0.066 | 35.9±5.2 |
| KIKINet | 1.9M | 0.960±0.020 | 34.9±0.2 | 0.891±0.033 | 34.6±1.9 | 0.889±0.074 | 32.1±4.4 |
| LPDNet | 118k | 0.976±0.007 | 37.2±0.0 | 0.907±0.028 | 35.4±1.9 | 0.898±0.083 | 31.5±4.6 |
| PICS | | 0.912±0.028 | 33.9±0.4 | 0.814±0.025 | 33.8±3.7 | 0.856±0.160 | 31.8±10.0 |
| RIM | 94k | 0.980±0.008 | 39.0±0.7 | 0.914±0.027 | 36.0±2.3 | 0.941±0.063 | 36.0±5.2 |
| UNet | 1.9M | 0.928±0.022 | 28.1±4.5 | 0.894±0.033 | 34.2±2.2 | 0.865±0.087 | 30.3±4.8 |
| Zero-Filled | | 0.869±0.056 | 20.1±1.1 | 0.823±0.017 | 22.8±0.9 | 0.824±0.084 | 21.0±4.6 |

Table S2: SSIM & PSNR scores of all methods evaluated on the $T_1$-Brain dataset (third and fourth column), the $T_2$-Knee dataset (fifth and sixth column), and the FLAIR-Brain dataset (seventh, eighth, ninth and tenth column). For all datasets performance is reported for six times acceleration using Gaussian 2D undersampling. The second column reports the total number of trainable parameters for each model. Best performing models are highlighted in bold. Methods are sorted in alphabetical order.

| Method | Params | $T_1$-Brain Gaussian 2D 6x | | $T_2$-Knee Gaussian 2D 6x | | FLAIR-Brain Gaussian 2D 6x | |
|---|---|---|---|---|---|---|---|
| | | SSIM ↑ | PSNR ↑ | SSIM ↑ | PSNR ↑ | SSIM ↑ | PSNR ↑ |
| CascadeNet | 1.1M | 0.953±0.022 | 32.5±0.8 | 0.886±0.035 | 34.0±2.4 | 0.907±0.079 | 32.1±4.7 |
| CIRIM | 264k | **0.975±0.010** | **37.6±0.3** | **0.901±0.032** | **35.1±2.2** | **0.932±0.073** | **35.1±5.3** |
| E2EVN | 19.6M | 0.963±0.014 | 33.9±1.0 | **0.901±0.032** | **35.1±2.2** | 0.891±0.084 | 30.8±5.4 |
| IRIM | 53k | 0.974±0.011 | 37.2±0.6 | 0.894±0.034 | 34.7±2.4 | 0.921±0.079 | 34.3±5.0 |
| KIKINet | 1.9M | 0.948±0.030 | 33.5±0.4 | 0.868±0.039 | 33.5±2.0 | 0.856±0.094 | 30.3±4.6 |
| LPDNet | 118k | 0.971±0.010 | 36.3±0.3 | 0.893±0.031 | 34.6±1.7 | 0.883±0.093 | 30.9±4.5 |
| PICS | | 0.889±0.029 | 32.4±0.5 | 0.779±0.031 | 32.1±3.9 | 0.842±0.163 | 31.0±8.9 |
| RIM | 94k | 0.974±0.010 | 37.5±0.7 | 0.896±0.033 | 34.5±2.3 | 0.926±0.075 | 34.6±5.2 |
| UNet | 1.9M | 0.910±0.035 | 27.8±3.6 | 0.872±0.040 | 32.8±2.4 | 0.829±0.102 | 28.7±4.7 |
| Zero-Filled | | 0.821±0.077 | 18.2±2.1 | 0.746±0.024 | 19.6±1.2 | 0.739±0.109 | 17.8±4.5 |

Table S3: SSIM & PSNR scores of all methods evaluated on the $T_1$-Brain dataset (third and fourth column), the $T_2$-Knee dataset (fifth and sixth column), and the FLAIR-Brain dataset (seventh, eighth, ninth and tenth column). For all datasets performance is reported for eight times acceleration using Gaussian 2D undersampling. The second column reports the total number of trainable parameters for each model. Best performing models are highlighted in bold. Methods are sorted in alphabetical order.

| Method | Params | $T_1$-Brain Gaussian 2D 8x | | $T_2$-Knee Gaussian 2D 8x | | FLAIR-Brain Gaussian 2D 8x | |
|---|---|---|---|---|---|---|---|
| | | SSIM ↑ | PSNR ↑ | SSIM ↑ | PSNR ↑ | SSIM ↑ | PSNR ↑ |
| CascadeNet | 1.1M | 0.940±0.031 | 31.8±0.8 | 0.870±0.040 | 33.0±2.5 | 0.888±0.090 | 31.1±4.6 |
| CIRIM | 264k | **0.970±0.012** | **36.6±0.5** | **0.888±0.043** | **34.3±2.3** | **0.922±0.082** | **34.2±5.1** |
| E2EVN | 19.6M | 0.952±0.020 | 32.6±1.3 | 0.887±0.036 | **34.3±2.5** | 0.870±0.094 | 30.0±4.8 |
| IRIM | 53k | 0.968±0.014 | 36.2±0.4 | 0.881±0.038 | 34.0±2.2 | 0.908±0.088 | 33.3±4.9 |
| KIKINet | 1.9M | 0.936±0.034 | 32.2±0.9 | 0.853±0.042 | 32.7±1.8 | 0.833±0.108 | 29.0±4.7 |
| LPDNet | 118k | 0.966±0.013 | 35.4±0.2 | 0.882±0.035 | 33.9±1.8 | 0.868±0.102 | 30.2±4.6 |
| PICS | | 0.875±0.030 | 31.5±0.7 | 0.752±0.036 | 30.7±4.0 | 0.834±0.164 | 30.5±8.2 |
| RIM | 94k | 0.969±0.013 | 36.3±0.5 | 0.883±0.036 | 34.1±2.2 | 0.914±0.084 | 33.6±5.0 |
| UNet | 1.9M | 0.891±0.042 | 27.1±3.3 | 0.857±0.044 | 32.1±2.1 | 0.800±0.114 | 27.6±4.5 |
| Zero-Filled | | 0.790±0.080 | 17.7±2.0 | 0.702±0.029 | 18.2±1.2 | 0.688±0.123 | 16.5±4.6 |

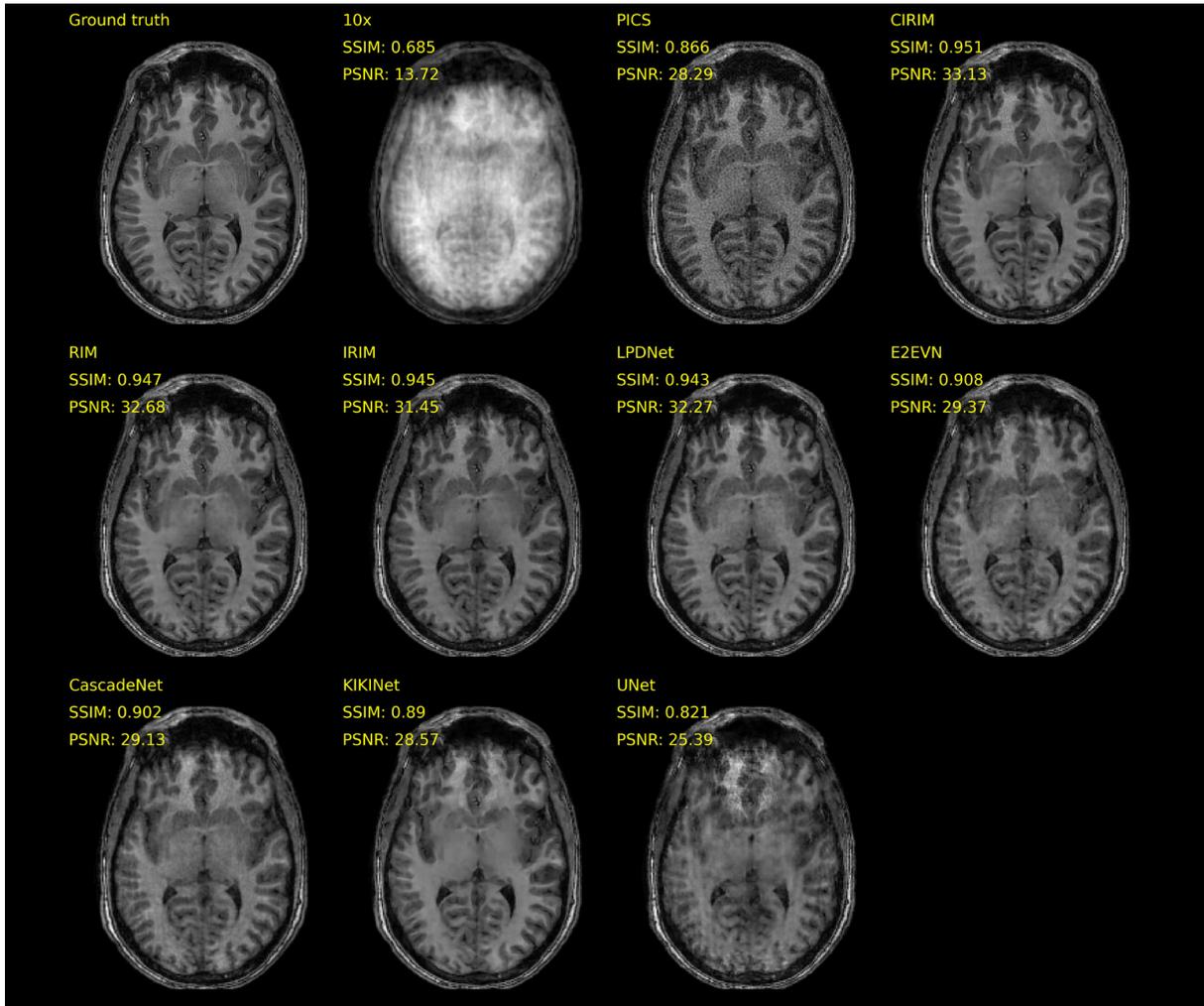

Figure S3: Reconstructions of a ten times accelerated slice with a Gaussian 2D mask, from the validation set of the $T_1$-weighted brains dataset (first row-second). The ground truth is presented on the first row-first image. The CIRIM 7C (first row-fourth), the RIM (second row-first), and the IRIM (second row-second) enforced Data Consistency (DC) implicitly by gradient descent. The E2EVN 8C (second row-fourth), the CascadeNet (third row-first), and the KIKINet (third row-second) enforced DC explicitly by a formulated DC term.

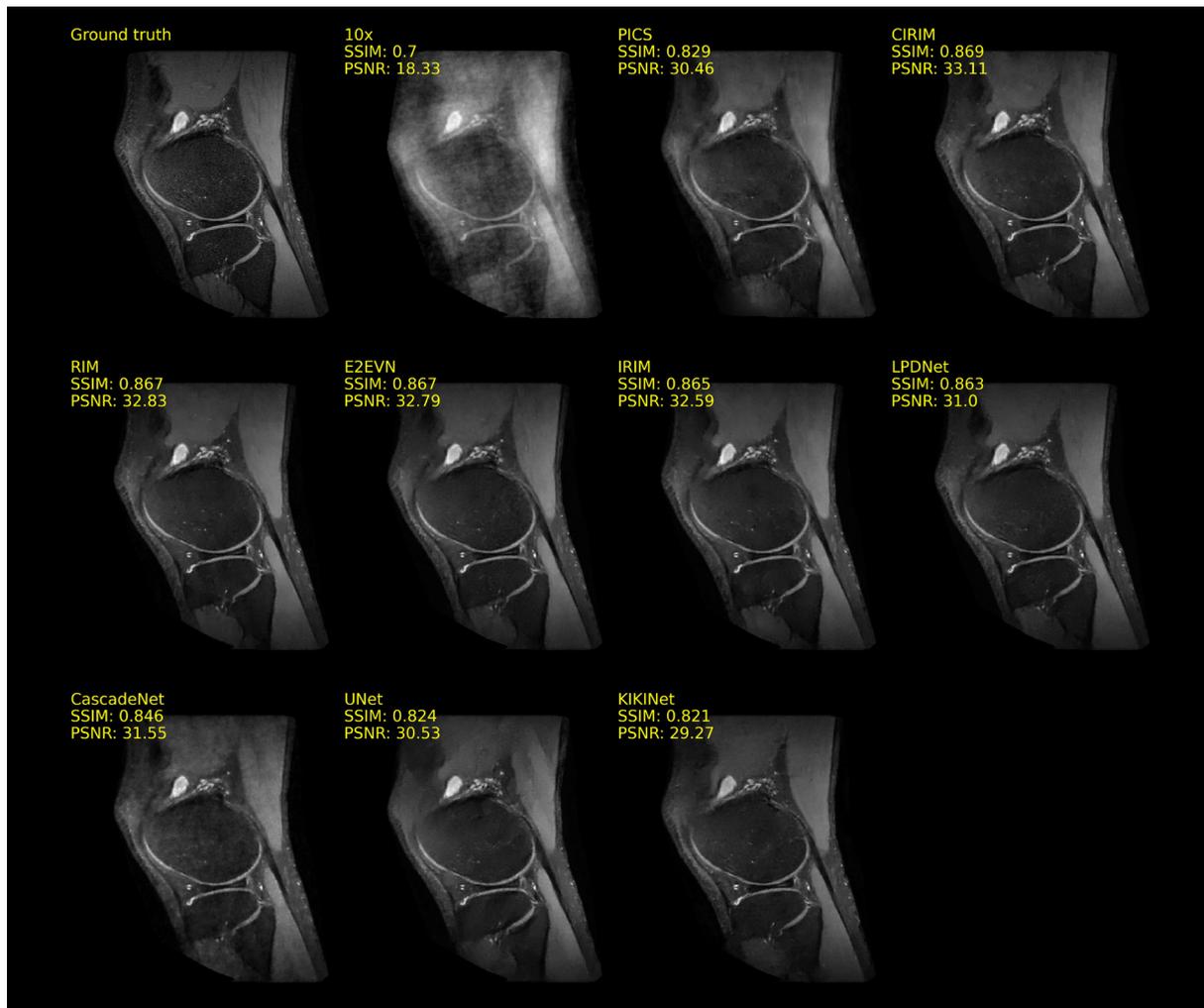

Figure S4: Reconstructions of a ten times accelerated slice with a Gaussian 2D mask, from the validation set of the $T_2$-weighted knees dataset (first row-second). The ground truth is presented on the first row-first image. The CIRIM 5C (first row-fourth), the RIM (second row-first), and the IRIM (second row-third) enforced Data Consistency (DC) implicitly by gradient descent. The E2EVN 8C (second row-second), the CascadeNet (third row-first), and the KIKINet (third row-third) enforced DC explicitly by a formulated DC term.

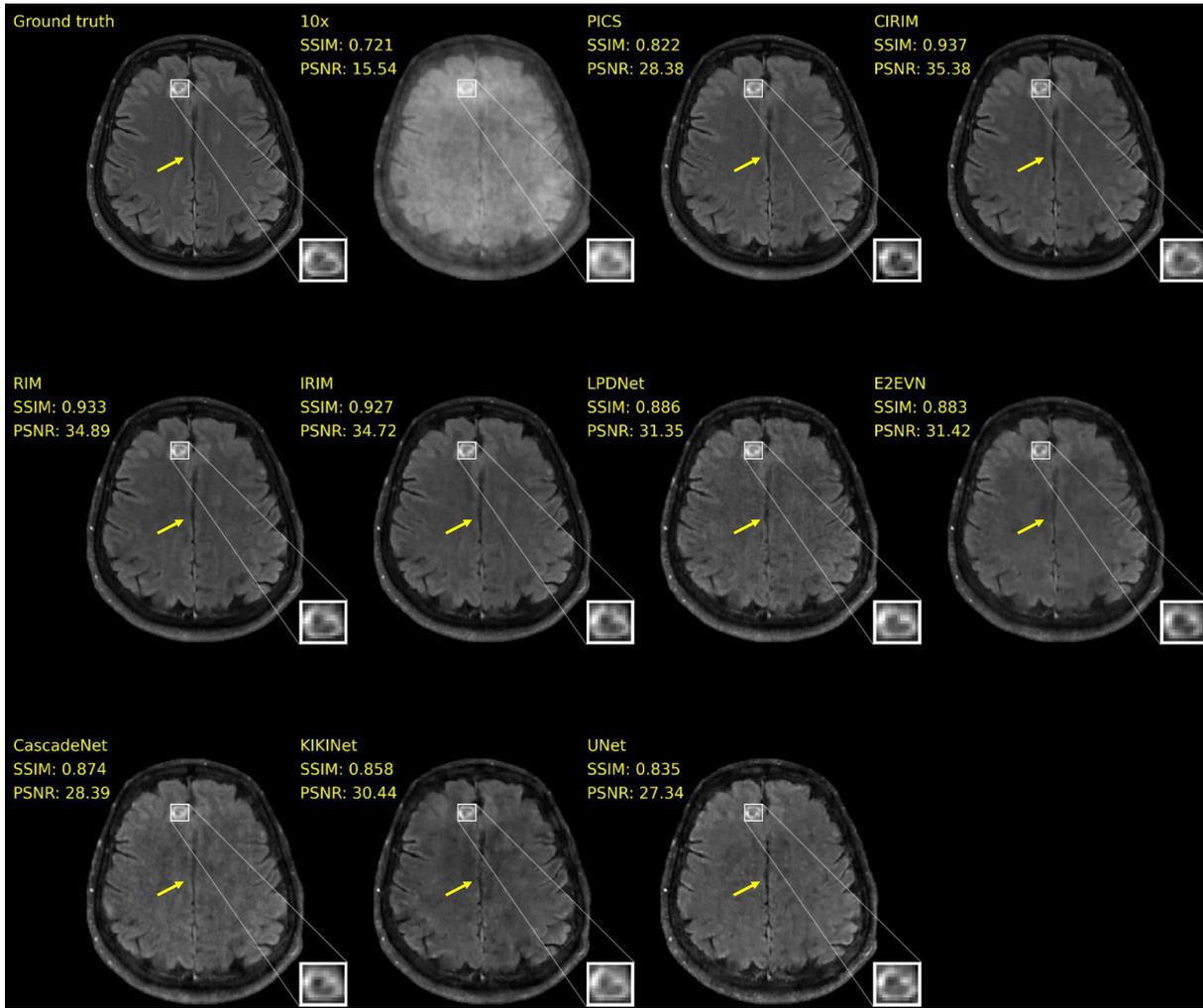

Figure S5: Reconstructions of a ten times accelerated slice with a Gaussian 2D mask, from the validation set of the FLAIR brains dataset (first row-second). The ground truth is presented on the first row-first image. The inset focuses on a reconstructed White Matter lesion; obtained through the fastMRI+ annotations (Zhao et al., 2021). The arrow points out to a region of interested that some models failed to reconstruct. The CIRIM 5C (first row-fourth), the RIM (second row-first), and the IRIM (second row-second) enforced Data Consistency (DC) implicitly by gradient descent. The E2EVN 8C (second row-fourth), the CascadeNet (third row-first), and the KIKINet (third row-second) enforced DC explicitly by a formulated DC term.

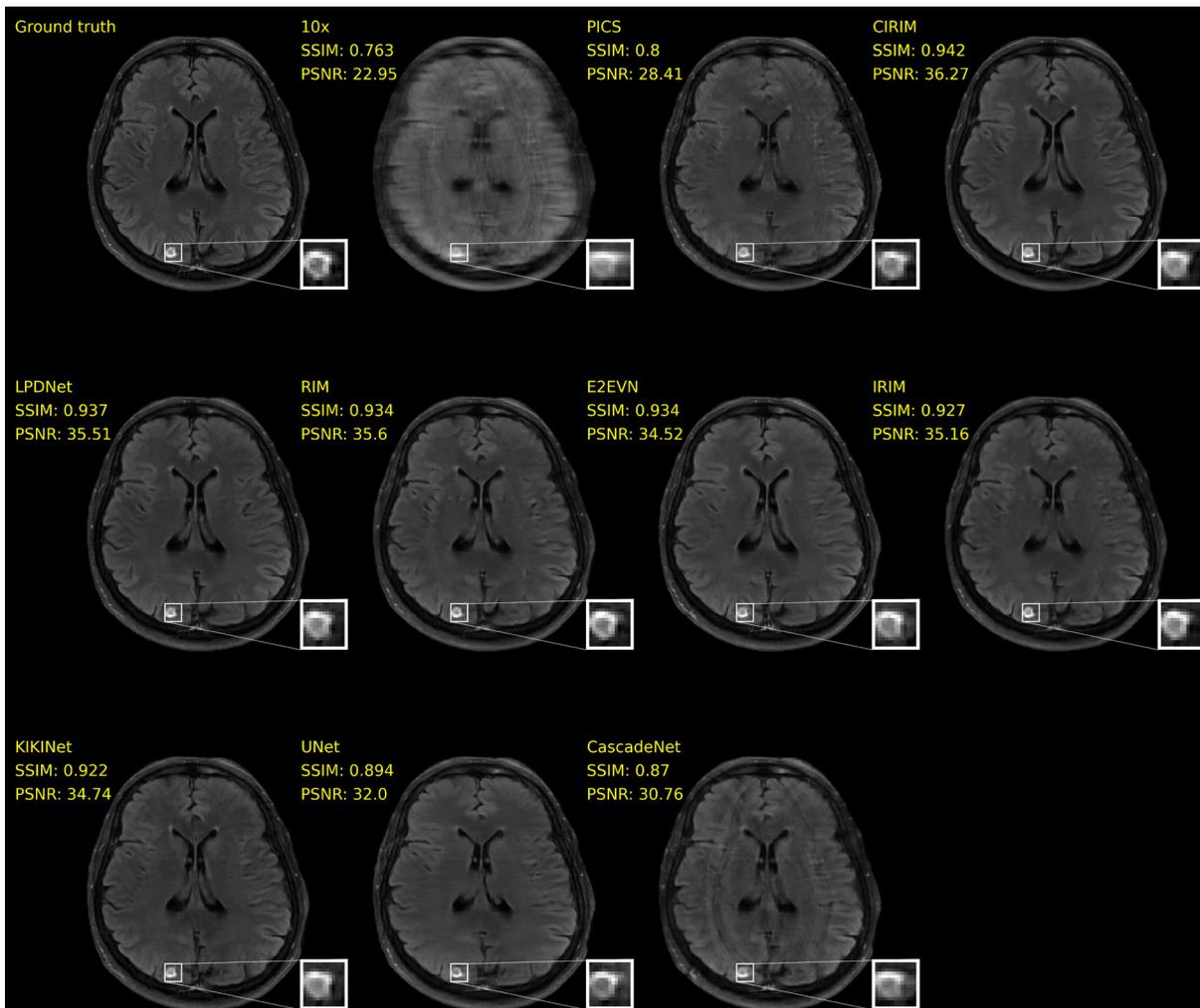

Figure S6: Reconstructions of a four times accelerated slice with an equidistant 1D mask, from the validation set of the FLAIR brains dataset (first row-second). The ground truth is presented on the first row-first image. The inset focuses on a reconstructed White Matter lesion; obtained through the fastMRI+ annotations (Zhao et al., 2021). The CIRIM 5C (first row-fourth), the RIM (second row- second), and the IRIM (second row- fourth) enforced Data Consistency (DC) implicitly by gradient descent. The E2EVN 8C (second row-third), the KIKINet (third row-first), and the CascadeNet (third row-third) enforced DC explicitly by a formulated DC term.

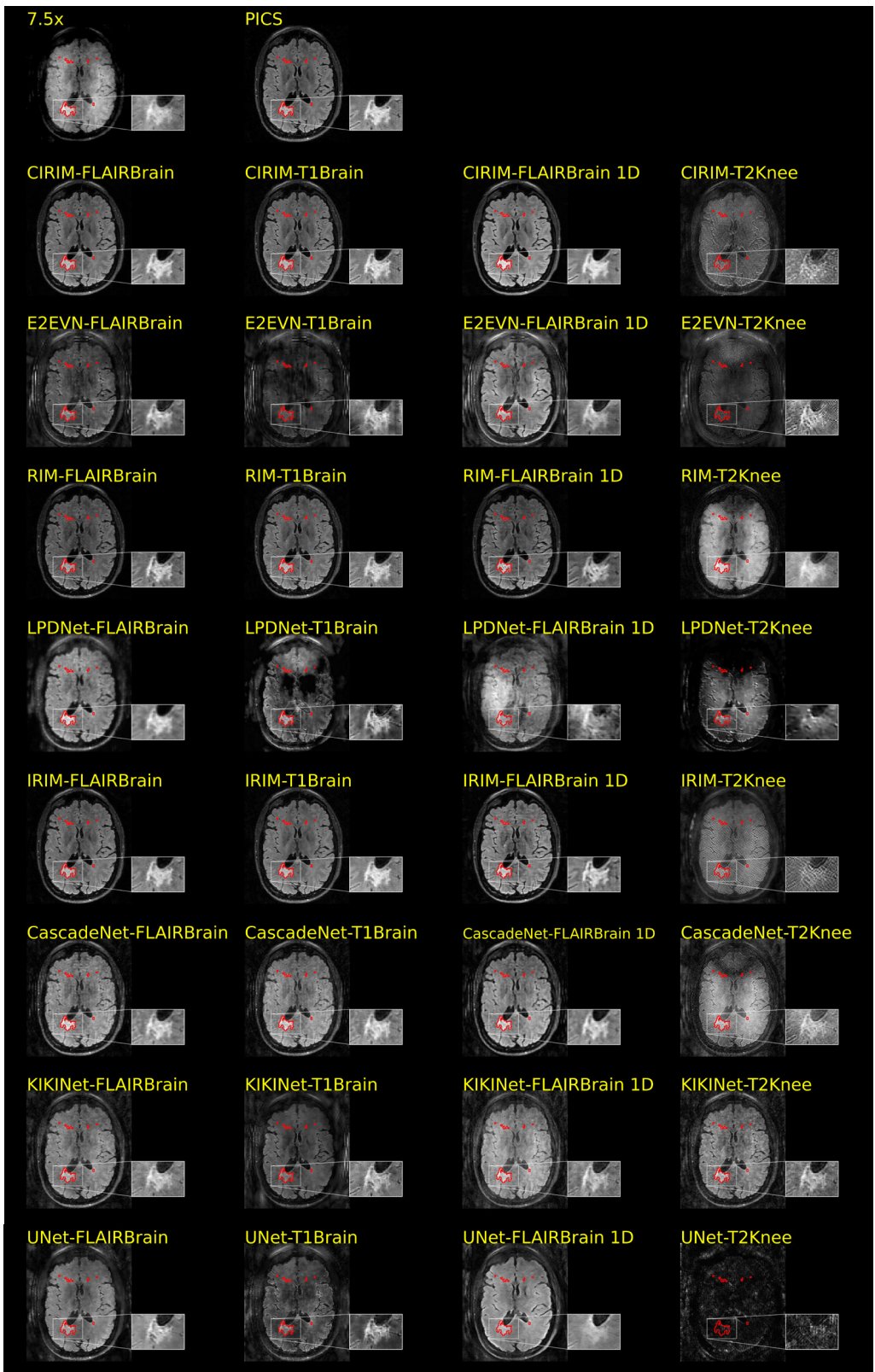

Figure S7: Reconstructions of a representative axial slice of a 7.5x accelerated 3D FLAIR scan of a MS patient. Segmented MS lesions are depicted with red colored contours. Shown is the aliased linear reconstruction (first row-first image), PICS (first row-second image), and models' reconstructions on each trained scheme: the FLAIR-Brain dataset with Gaussian 2D undersampling (second-last row, first column), the $T_1$-Brain dataset with Gaussian 2D undersampling (second-last row, second column), the FLAIR-Brain dataset with equidistant 1D undersampling (second-last row, third column), and the $T_2$-Knee dataset with Gaussian 2D undersampling (second-last row, fourth column). The inset on the right bottom of each reconstruction focuses on a lesion region with high spatial detail.

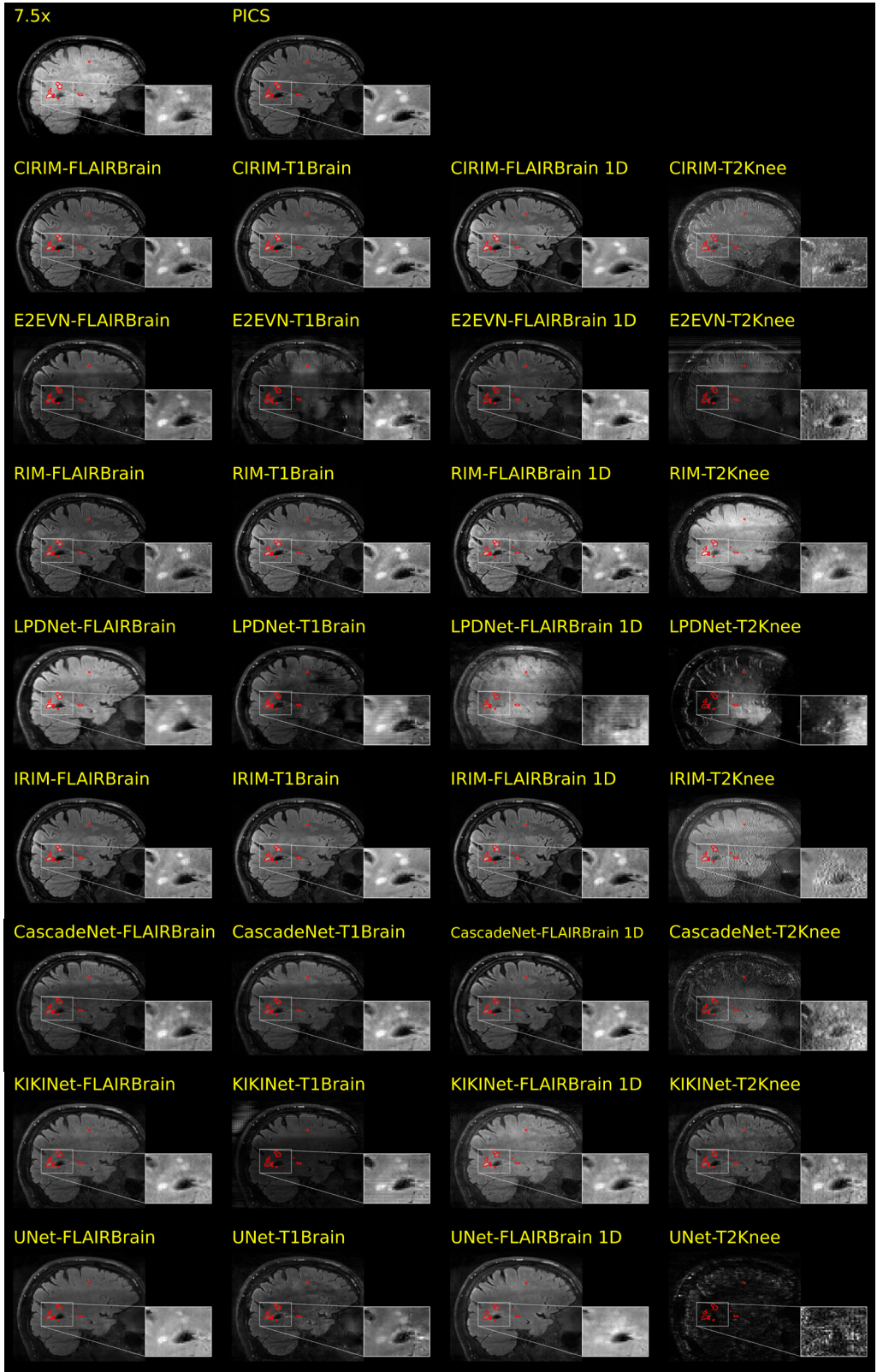

Figure S8: Reconstructions of a representative sagittal slice of a 7.5x accelerated 3D FLAIR scan of a MS patient. Segmented MS lesions are depicted with red colored contours. Shown is the aliased linear reconstruction (first row-first image), PICS (first row-second image), and models' reconstructions on each trained scheme: the FLAIR-Brain dataset with Gaussian 2D undersampling (second-last row, first column), the $T_1$-Brain dataset with Gaussian 2D undersampling (second-last row, second column), the FLAIR-Brain dataset with equidistant 1D undersampling (second-last row, third column), and the $T_2$-Knee dataset with Gaussian 2D undersampling (second-last row, fourth column). The inset on the right bottom of each reconstruction focuses on a lesion region with high spatial detail.